\documentclass[10pt, twocolumn]{IEEEtran}
\usepackage[left=1.57cm,right=1.57cm,top=1.52cm,bottom=3cm]{geometry}

\usepackage{graphicx,cite,acronym,amsmath,amssymb,amsfonts,color,floatflt,stfloats}
\usepackage{caption}
\usepackage{subcaption}
\usepackage{epsfig,epstopdf,psfrag}
\usepackage[ruled,vlined]{algorithm2e}
\SetKwInput{KwInput}{Input}                
\SetKwInput{KwOutput}{Output}  
\usepackage[table]{xcolor}
\usepackage{bm}
\usepackage[cmintegrals]{newtxmath}
\usepackage{graphics,hhline,array,cite,bbm}
\usepackage{latexsym}
\usepackage{theorem}
\usepackage{times}
\usepackage{makeidx}
\usepackage{bbm}
\usepackage{tikz}
\usepackage{amsmath}
\def\BibTeX{\rm B\kern-.05em{\sc i\kern-.025em b}\kern-.08em
    T\kern-.1667em\lower.7ex\hbox{E}\kern-.125emX}
\usepackage{hyperref}

\acrodef{5G}{fifth generation}
\acrodef{6G}{sixth generation}
\acrodef{AOA}{angle-of-arrival}
\acrodef{AOD}{angle-of-departure}
\acrodef{AI}{artificial intelligence}
\acrodef{UAV-I}{{\em UAV intelligence}}
\acrodef{AI/ML}{artificial intelligence/machine learning}
\acrodef{A2A}{agent-to-agent}
\acrodef{AWGN}{additive white Gaussian noise}
\acrodef{AO}{alternating optimization}
\acrodef{AP}{Access Point}
\acrodef{BS}{base station}
\acrodef{BCD}{Block Coordinate Descent}
\acrodef{BD}{beyond diagonal}
\acrodef{CRLB}{Cram\'er-Rao Lower Bound}
\acrodef{CDF}{cumulative density function}
\acrodef{CDR}{correct detection rate}
\acrodef{CIR}{channel impulse response}
\acrodef{CRAS}{connected robotics and autonomous systems}
\acrodef{CNPC}{control and non-payload communication}
\acrodef{CSI}{Channel State Information}
\acrodef{CR}{Constraint Relaxation}
\acrodef{DEC-POMDP}{decentralized partially observable Markov decision process}
\acrodef{D2D}{device-to-device}
\acrodef{DC}{dual connectivity}
\acrodef{DL}{downlink}
\acrodef{EKF}{extended Kalman filter}
\acrodef{EIRP}{effective isotropic radiated power}
\acrodef{FAR}{false alarm rate}
\acrodef{FIM}{Fisher Information matrix}
\acrodef{FL}{federated learning}
\acrodef{FMCW}{frequency modulated continuous wave}
\acrodef{FDMA}{Frequency Division Multiple Access}
\acrodef{FD}{full duplex}
\acrodef{GLRT}{generalized likelihood ratio test}
\acrodef{GP}{Gaussian process}
\acrodef{GPI}{generalized policy iteration}
\acrodef{GAM}{gradient ascent method}
\acrodef{IoT}{Internet-of-Things}
\acrodef{IS}{image similarity}
\acrodef{IRS}{intelligent reflecting surfaces}
\acrodef{IO}{indoor office}
\acrodef{ISI}{inter-symbol interference}
\acrodef{KPI}{key performance indicator}
\acrodef{KKT}{Karush Kuhn Tucker}
\acrodef{KF}{Kalman filter}
\acrodef{LLRT}{log-likelihood ratio test}
\acrodef{LOS}{line-of-sight}
 \acrodef{MAC}{medium access control}
\acrodef{MAP}{maximum a-posteriori probability}
\acrodef{MEC}{multi-access edge computing}
\acrodef{MAB}{multi-armed bandit}
\acrodef{mMTC}{massive machine type communication}
\acrodef{eMBB}{enhanced mobile broadband}
\acrodef{MMSE}{minimum mean squared error }
\acrodef{MARL}{multi-agent reinforcement learning}
\acrodef{MDP}{Markov decision process}
\acrodef{MIMO}{multiple-input multiple-output}
\acrodef{ML}{machine learning}
\acrodef{MLE}{maximum likelihood estimator}
\acrodef{mm-wave}{millimeter-wave}
\acrodef{MISO}{multiple input single output}
\acrodef{MSE}{mean square error}
\acrodef{MRT}{Maximum Ratio Transmission}
\acrodef{MM}{Minorization-Maximization}
\acrodef{NOMA}{non-orthogonal multiple access}
\acrodef{NLOS}{non line-of-sight}
\acrodef{OG}{occupancy grid}
\acrodef{OFDM}{orthogonal frequency division multiplexing}
\acrodef{OFDMA}{orthogonal frequency division multiple access}
\acrodef{PF}{particle filtering}
\acrodef{pdf}{probability density function}
\acrodef{PFA}{probability of false alarm}
\acrodef{PL}{packet loss}
\acrodef{POMDP}{partially observable Markov decision process}
\acrodef{PER}{packet error rate}
\acrodef{PEB}{position error bound}
\acrodef{PD}{Path Discretization}
\acrodef{QoS}{quality of service}
\acrodef{RIS}{reconfigurable intelligent surface}
\acrodef{RCS}{radar cross section}
\acrodef{RMSE}{root mean square error}
\acrodef{RFID}{radiofrequency identification}
\acrodef{RL}{reinforcement learning}
\acrodef{ROC}{receiver operating characteristics}
\acrodef{RR}{Round Robin}
\acrodef{RRCS}{root-radar cross section}
\acrodef{RV}{random variable}
\acrodef{ROI}{region of interest}
\acrodef{SLAM}{simultaneous localization and mapping}
\acrodef{SNR}{signal-to-noise ratio}
\acrodef{SIR}{sequential importance resampling} 
\acrodef{SOCP}{second-order cone programming}
\acrodef{SDR}{semidefinite relaxation}
\acrodef{SM}{shopping mall}
\acrodef{SINR}{signal-to-interference-plus-noise ratio}
\acrodef{SE}{Spectral Efficiency}
\acrodef{SCA}{successive convex approximation}
\acrodef{STAR}{simultaneously transmitting and reflecting}
\acrodef{SIC}{successive interference cancelation}
\acrodef{SDP}{semidefinite programming}
\acrodef{SISO}{Single Input Single Output}
\acrodef{TD}{temporal-difference}
\acrodef{TOA}{time-of-arrival}
\acrodef{TDMA}{Time Division Multiple Access}
\acrodef{UE}{user equipment}
\acrodef{THz}{Terahertz}
\acrodef{UAV}{unmanned aerial vehicle}
\acrodef{U2U}{UAV-to-UAV}
\acrodef{UKF}{Unscented Kalman Filter}
\acrodef{UCB}{upper confidence bound}
\acrodef{URLLC}{ultrareliable and low-latency communication}
\acrodef{UL}{uplink}
\acrodef{VLC}{visible light communication}

\acrodef{WMMSE}{weighted Minimum Mean Square Error}
\acrodef{SOCP}{second order cone programming}
\acrodef{LC}{local constraint}
\acrodef{GC}{global constraint}
\acrodef{LDP}{Lagrangian dual problem}
\usepackage{comment}
\usepackage{tikz}
\usepackage{pgfplots}
\usepackage[ruled,vlined]{algorithm2e}
\usepackage{psfrag}
\usepackage{soul}
\usepackage{graphicx}
\usepackage{textcomp}
\usepackage[table]{xcolor}
\usepackage{bbm}
\usepackage{caption}
\usepackage{subcaption}
 
\usepackage{amsmath,amssymb,amsfonts, acronym}
 
\usepackage{hyperref}
\usepackage{xparse}
\usepackage{float}
\usepackage{pstricks}
\usepackage{amsthm}
\usepackage[squaren,Gray]{SIunits}
\usepackage{soul}

\usepackage[singlespacing]{setspace} 

\definecolor{silver}{rgb}{0.95, 0.95, 0.95}
\definecolor{darkgreen}{rgb}{0.0, 0.5, 0}
\definecolor{JungleGreen}{rgb}{0.16, 0.67, 0.53}

\usetikzlibrary{positioning} 
\definecolor{Eored}{rgb}{.647 ,.129 ,.149} 
\definecolor{Eogreen}{rgb}{0 ,0.53 ,0}

\newcommand{\mbf}{\mathbf}

\newcommand{\Gt} {G_{\text{T}}}
\newcommand{\Gr} {G_{\text{R}}}

\DeclareMathOperator{\E}{\mathbb{E}}

\newcommand{\txsymbol}{\theta}

\begin{document}
 
\author{Silvia~Palmucci \IEEEmembership{Student~Member,~IEEE}, Giulio~Bartoli \IEEEmembership{Member,~IEEE},\\ Andrea~Abrardo \IEEEmembership{Senior~Member,~IEEE}, Marco~Moretti \IEEEmembership{Member,~IEEE} \\ Marco~Di~Renzo \IEEEmembership{Fellow~Member,~IEEE}

\thanks{S. Palmucci, G. Bartoli and A. Abrardo are with the University of Siena and CNIT, Italy (e-mail: silvia.palmucci@student.unisi.it, giulio.bartoli@unisi.it, abrardo@dii.unisi.it). M. Moretti is with Dipartimento di Ingegneria dell'Informazione, University of Pisa and CNIT, Italy (e-mail: marco.moretti@unipi.it). M. Di Renzo is with Universit\'e Paris-Saclay, CNRS, CentraleSup\'elec, Laboratoire des Signaux et Syst\`emes, 3 Rue Joliot-Curie, 91192 Gif-sur-Yvette, France. (marco.di-renzo@universite-paris-saclay.fr).
The work of S. Palmucci and A. Abrardo is supported by the 6G SHINE european project.
The work of G. Bartoli is supported by the PSR 2023 - New Frontiers and by the European fund FSE REACT-EU, PON Ricerca e Innovazione 2014-2020.
The work of M. Moretti is partially supported by the MIUR in the FoReLab project.
The work of M. Di Renzo was supported in part by the European Commission through the Horizon Europe project titled COVER under grant agreement number 101086228, the Horizon Europe project titled UNITE under grant agreement number 101129618, and the Horizon Europe project titled INSTINCT under grant agreement number 101139161, as well as by the Agence Nationale de la Recherche (ANR) through the France 2030 project titled ANR-PEPR Networks of the Future under grant agreement NF-YACARI 22-PEFT-0005, and by the CHIST-ERA project titled PASSIONATE under grant agreements CHIST-ERA-22-WAI-04 and ANR-23-CHR4-0003-01.
}
}
\title{{Power Minimization with Rate Constraints for Multi-User MIMO Systems with Large-Size RISs}}

\maketitle
\begin{abstract}
This study focuses on the optimization of a single-cell multi-user {\ac{MIMO}} system with multiple {large-size} \acp{RIS}. The overall transmit power is minimized by optimizing the precoding coefficients and the RIS configuration, with constraints on users' \acp{SINR}.
The minimization problem is divided into two sub-problems and solved by means of an iterative \ac{AO} approach. 
The first sub-problem focuses on finding the best precoder design.
 The second sub-problem optimizes the configuration of the RISs by partitioning them into smaller tiles. Each tile is then configured as a combination of pre-defined {configurations.}
 This allows the {efficient optimization of RISs}, especially in scenarios where the computational complexity would be prohibitive using traditional approaches.
Simulation results show the good performance and limited complexity of the proposed method in comparison {to} benchmark {schemes}.

\end{abstract}
\acresetall

\section{Introduction}

{Recent advancements in \acp{RIS} have revolutionized wireless communication systems by enabling dynamic control over signal propagation environments. RIS, composed of numerous reflecting elements, can effectively manipulate electromagnetic waves, offering unprecedented flexibility in signal enhancement and interference mitigation in the age of 5G and beyond \cite{direnzo2020smart}. 
 Traditionally, RISs are seen as a tool to fill coverage gaps, enabling the service of 'unlucky' nodes that lack adequate network coverage or, due to their quasi-passive nature, as a mean to minimize energy consumption while meeting certain communication quality constraints. Specifically, power minimization, considered a simpler and special case of the energy maximization problem \cite{fotock2024energy}, is a fundamental objective in wireless networks to prolong the battery life of the devices and reduce operational costs. By judiciously adjusting the reflection coefficients of RIS elements, it becomes feasible to optimize signal strength and quality while minimizing overall power consumption.
The combination of power minimization, {with }achievable rate constraints finds applications in many practical scenarios. Examples include green communications, industrial Internet of Things, indoor wireless networks, and emergency systems, where it is preferable or mandatory to serve a large number of nodes while ensuring proportional benefits to all.}

\subsection{State-of-the-art}
{In the following we restrain our attention to working scenarios with nearly-passive devices, that solely reflect the impinging signals avoiding the use of power amplifiers.}
{Minimizing the total transmit power at the \ac{BS}, while simultaneously ensuring that users within RIS-aided systems meet minimum rate constraints, is a problem that, to the best of our knowledge, has been studied in a relatively few works, e.g., \cite{wu2019intelligent, wu2020transmit, wu2021power, wang2022transmit, ma2022transmit, zhoy2023transmit, feng2023resource, ren2023energy, xie2021joint, sun2024new, kumar2023novel, kai2024delay}.}
In one of the pioneering works in this field, \cite{wu2019intelligent}, the authors consider a multi-user MIMO system and formulate the minimization problem as a joint optimization of the transmit beamforming at the \ac{BS} and the reflection coefficients at the \ac{RIS}. By employing an iterative \ac{AO} approach, the {formulated} optimization problem is decomposed into subproblems that are solved by using convex optimization tools such as \ac{SOCP} and \ac{SDR}. Building upon this work, several other studies have been undertaken 
{ with alternative approaches. In \cite{wu2020transmit} and \cite{wu2021power} the optimization problem is solved resorting a manifold-based \ac{AO} algorithm.
{In}\cite{wang2022transmit,ma2022transmit,zhoy2023transmit} in which a \ac{STAR}-RIS-aided communication system is studied, the optimization problem is solved by means of \ac{SCA} and \ac{SDP}.
Some examples in \ac{NOMA} scenarios have been reported in \cite{xie2021joint, feng2023resource, ren2023energy}, with \cite{feng2023resource, ren2023energy} specifically addressing \ac{UAV}-assisted environments. 
A \ac{BD} RIS scenario is analyzed in \cite{sun2024new}, in which RIS optimization is performed using Takagi factorization. 
In a recent paper \cite{kumar2023novel}, the authors propose a new approach to solve the power minimization problem with \ac{QoS} constraints, as formulated in \cite{wu2019intelligent}. Specifically, the paper introduces an SCA-based \ac{SOCP} algorithm in which all optimization variables are updated simultaneously in each iteration, thereby eliminating the need for AO. The approach has been applied in different other works, e.g., \cite{kai2024delay}.}

A relevant aspect that must be considered in all the cited works, and in general in all works dealing with RISs, is the computational complexity. Generally speaking, the algorithmic complexity of the proposed solutions is primarily rooted in two aspects: the { \emph{non-convexity}, inherently related to the presence of RISs, and} the \emph{high dimensionality} of the considered optimization problems. 
{As for the latter aspect,} large-size RISs are usually required in practical network deployments, in order to attain the necessary beamforming gain for coverage enhancement \cite{sihlbom2022reconfigurable, yuan2024reconfigurable }. Large-size RISs are inherently constituted by a large number of reconfigurable scattering elements, also referred as unit cells, each of which being an optimization variable. Therefore, the number of variables in the RIS optimization problem tends to be very large, and the computational complexity increases accordingly. For instance, in existing works, the iterative algorithms employed for optimization exhibit a complexity that scales at least with the cube of the total number of RIS elements \cite{wu2019intelligent, wang2022transmit, ma2022transmit, zhoy2023transmit, feng2023resource, ren2023energy, kumar2023novel}. 
In order to address the high dimensionality of the considered problem, it is possible to leverage RIS designs that group unit cells together. 
The concept of dividing an RIS into groups of unit cells, referred to as tiles, super cells, or cell groups, has already been considered in prior research. For example, in designing an RIS to redirect a plane wave to a non-specular direction in the far field, the reflection coefficient must be a periodic function, with the RIS resulting in a periodic repetition of a fundamental structure, the super cell. This is discussed in \cite{direnzo2020smart} and references therein, where a metasurface is defined as a periodic arrangement of unit cells, in non-uniform patterns, to simplify the design.
In complex scenarios with many network nodes and multipath, the optimization of RISs cannot be cast in a simple beam redirection problem and the periodicity assumption does not usually hold anymore. Each scattering element should be independently optimized to maximize the beamforming gain, significantly increasing the design complexity. To address this issue, adjacent cells can be designed to share an optimized state, reducing the optimization complexity but the limiting design flexibility, as proposed in \cite{rafique2023reconfigurable}.
To further reduce the number of possible states, predefined configurations, or codebooks, have been introduced. These methods quantize the possible angles of arrival and departure into discrete sets, as elaborated in \cite{albanese2022marisa} for beam sweeping and channel estimation, and in \cite{9306896} to reduce the optimization complexity under ideal channel estimation assumption. However, this approach reduces the design flexibility and is less effective in multipath environments with multiple nodes.
To overcome this limitation, \cite{Abrardo2021} suggests dividing the RIS into independently optimized tiles using an AO method. While this allows for a greater design flexibility, it also introduces high complexity, as no predefined configurations are used.

\subsection{Main Contributions}
\label{sec:main_contributions}
{ The main goal of this work is to propose a low-complexity solution to the problem of transmit power minimization with rate constraints in a multi-user \ac{MIMO} system by optimizing the \ac{BS} precoder and \ac{RIS} design. The problem is addressed in scenarios consisting of a large number of \ac{RIS} elements and nodes. In comparison to the state of the art, the main contributions of this paper can be summarized in two key points: the reduction of the dimensionality of the problem to be solved and the proposed strategy to reformulate the formulated non-convex problem into a convex one. Consequently, the key contribution is a significant reduction of the computational complexity compared to existing methods, while ensuring comparable performance.
Next, we elaborate these aspects in detail.}

\subsubsection*{Reduction of the Problem Dimensionality}
{
In this work, we build upon the concept of tiles proposed in \cite{Abrardo2021} to address a power minimization problem subject to rate constraints. To reduce the complexity, we adopt the idea of defining a set of possible base configurations for each tile {in order to recast the problem of optimizing a single RIS element into a problem of tile design.} To limit the loss in terms of design flexibility and to minimize the dimensionality of the optimization problem, instead of using fixed codebooks as in \cite{albanese2022marisa} and \cite{9306896}, the base functions are designed to maximize the power of the signal reflected from the tile towards each potential user, assuming perfect channel knowledge. This approach achieves performance comparable to available benchmark schemes that offer the highest flexibility by treating each RIS element as an independent variable, while significantly reducing the dimensionality of the problem.}

\subsubsection*{Convex Reformulation of the Problem} 
As in \cite{wu2019intelligent}, the optimization of the precoder is tackled by using the \ac{SOCP} method, and by subsequently optimizing the RIS configuration. However, unlike \cite{wu2019intelligent}, the RIS configuration is not obtained by applying the \ac{SDR} method and Gaussian randomization. Instead, we reformulate the (tiled) RIS configuration problem as an \ac{MMSE} problem with convex constraints. 

\subsubsection*{Reduction of the Computational Complexity}
{Given the formulated \ac{MMSE} problem, we propose a low-complexity solution using an optimization algorithm that operates in the dual domain, where the Lagrangian dual function and its gradient are computed in closed form. This enables the problem to be solved iteratively through the \ac{GAM}. We evaluate the complexity of the proposed algorithm and prove that is be significantly lower than that of well-known benchmark approaches available in the literature. The reduction in computational complexity  is shown to become more pronounced as the size of the RIS increases, making the proposed approach particularly attractive in practical scenarios where large-size RISs need to be deployed.}

{The performance of the proposed approach is assessed by comparing it against two main benchmarks proposed in \cite{wu2019intelligent} and \cite{kumar2023novel}}. 
The obtained numerical results highlight that the proposed simplified approach provides comparable results with respect to (w.r.t.) the typical and computationally demanding SDR approach \cite{wu2019intelligent}. {Better results are also obtained in terms of the \ac{SCA} method proposed in \cite{kumar2023novel}.}

\subsection{Paper Organization and Notation}
The rest of the paper is organized as follows: Section \ref{sec:model} introduces the system model, Sections \ref{sec:probelm_formulation}-\ref{optSolution} formulate the optimization problem and present the proposed solution, Section \ref{sec:results} shows several numerical results and, Section \ref{sec:future} draws some final conclusions.\\
 
\textit{Notation}: Scalar variables, vectors, and matrices are represented with lower letters, lower bold letters, and capital bold letters, respectively (e.g., $x$, $\mathbf{x}$, and $\mathbf{X}$). The symbols $\left( \cdot \right)^H$ and $\left( \cdot \right)^{-1}$ represent the conjugate transpose and inverse operators of their arguments, respectively. The notation $x = (\cdot)^+$ assigns to x the maximum value between $0$ and the argument. $|| \cdot ||$ and $|\cdot|$ represent the norm and the absolute value of the argument. $\Delta_{x}$ and $\frac{\partial}{\partial_x}$ indicate the derivative of the argument w.r.t. the variable $x$. $\Re[\cdot]$ indicates the real part of the argument and $\angle{\cdot}$ is used to indicate the projection operation while $\measuredangle{\cdot}$ refers to the phase of the argument. Finally $\mathbb{E}[\cdot]$ indicates the expectation of the argument, $\mathcal{O}(\cdot)$ stands for the big-O notation, and $\otimes$ is the Kronecker product. 

\section{System Model}
\label{sec:model}
\begin{figure}
    \centering
    \includegraphics[width = 0.8\linewidth]{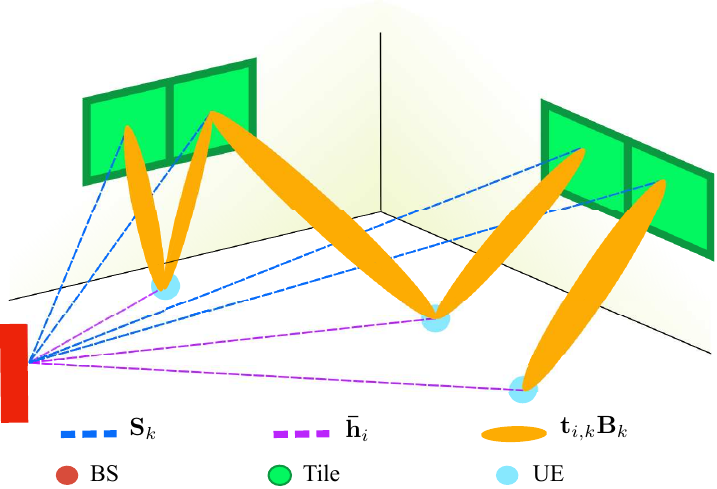}
    \caption{Example of system model for $Q = 2$, $C = 2$, $K = 4$, $N_u = 3$, with $i = 1, \ldots, N_u$, $k = 1, \ldots, K$.}
    \label{fig:scenarioTiles}
\end{figure}

We consider the downlink of a multi-user {MIMO} system aided by multiple RISs, where one \ac{BS} equipped with $M$ antennas serves $N_u$ single-antenna \acp{UE}{, as sketched} in Figure \ref{fig:scenarioTiles}. 
The total number of RISs is $Q$ and each \ac{RIS} consists of $C$ non overlapping tiles, so that the total number of tiles is $K = Q C$. Moreover, each tile is composed by $P$ scattering elements. 
To cope with the concurrent transmissions of several users, the \ac{BS} employs spatial precoding to enhance the transmission efficiency and to mitigate inter-user interference. The information symbol for the $j$-th \ac{UE}, denoted by $\txsymbol_j \in \mathbb{C}$, is modelled as a zero-mean \ac{RV} with unitary mean square value. {Different symbols are considered} independent and identically distributed (i.i.d.). The $j$-th transmitted data stream $\mathbf{x}_j\in \mathbb{C}^{M \times 1}$ is obtained by multiplying the information data by the corresponding precoding vector $\mathbf{v}_j\in \mathbb{C}^{M}$, i.e., 
\begin{equation}
\mathbf{x}_j = \mathbf{v}_j \txsymbol_j \, .
\end{equation}
At each of the $N_u$ receivers, the signal is the superimposition of the signals reflected by all the RISs in the system. Accordingly, the signal ${z}_{i,k,j} \in \mathbb{C}$, received at the $i$-th {UE}, reflected by the $k$-th tile, and relative to the $j$-th transmitted stream can be expressed as
\begin{equation}
{z}_{i,k,j} = \mathbf{t}_{i,k} \mathbf{B}_{k} \mathbf{S}_{k}\mathbf{x}_j,
\label{eq:rx_irs3}
\end{equation}
where $\mathbf{t}_{i,k} \in\mathbb{C}^{1 \times P}$ is the channel vector between the $k$-th tile and the $i$-th receiver, $\mathbf{B}_{k} = \text{diag}\left(\mathbf{b}_{k}\right) \in \mathbb{C}^{P \times P}$ is a diagonal matrix whose elements are $b_{k,p}$, which models the scattering coefficient of the $p$-th scattering element ($p \in \left \{ 1, \ldots, P\right \}$) of the $k$-th tile, and $\mathbf{S}_{k} \in\mathbb{C}^{P \times M}$ is the channel matrix between the BS and the $k$-th tile. 

{In this work, we put forth the idea that the vector $\mathbf{b}_{k}$ can be represented as a linear combination of an appropriate set of basis vectors. By selecting an optimal basis vector, the design of the system can be significantly simplified. Specifically, we assume that $\mathbf{b}_{k}$ can be decomposed using $N_u$ basis vectors, $\mathbf{b}_{k}^{(m)} \in \mathbb{C}^{P \times 1}$, with $m = 1, \ldots, N_u$ and the number of basis vectors is set equal to the number of users, i.e.,}
\begin{equation}
\label{eq: base expansion}
\mathbf{b}_{k} = \sum_{m=1}^{N_u}\alpha_{m,k} \mathbf{b}_{k}^{(m)}.
\end{equation}
As for the rationale behind the design of the basis vector, the rationale is that each element $\mathbf{b}_k^{(m)}$ represents a beam from tile $k$ to node $m$. The details of the strategy for deriving the basis vectors, and thus the physical meaning of beam, is elaborated next. The toy example depicted in Figure \ref{fig:scenarioTiles} shows {a case} of possible RIS configuration. The presence of a beam from tile $k$ to node $m$ indicates that the corresponding coefficient $\alpha_{m,k}$ is different from zero. 
A single tile may serve multiple users if there is more than one non-zero coefficient $\alpha_{m,k}$. 
Similarly, when the user $m$ is linked to two different tiles the corresponding coefficients are different from zero.

{Assuming that each RIS is a nearly-passive device}\footnote{We refer to an RIS as a nearly-passive device because its configurability requires the necessity of receiving data from a central controller and the need of a control circuit and electronic circuits to change the state of the tunable elements. However, no power amplifiers RF chains, and digital signal processing are needed.}, the {amplitude of the reflection coefficient} $b_{k,p}$ ($p \in \left \{ 1, \ldots, P\right \},k \in \left \{ 1, \ldots, K\right \}$) {
of each unit cell scatters an amount of power that is at most equal to the impinging
power, i.e.,
\begin{equation}
\left|{b}_{k,p}\right|^2 = \left|\sum_{m=1}^{N_u}\alpha_{m,k} {b}_{k,p}^{(m)}\right|^2 \leq 1,
\label{cons_LC}
\end{equation}
where the equality holds in absence of losses. This latter case corresponds to consider that each element of the RIS introduces a simple phase shift, i.e., $b_{k,p} = e^{j\phi_{k,p}}$. The constraint in \eqref{cons_LC} is referred to as \ac{LC}.
} By introducing the matrix $\mathbf{B}^{(m)}_{k} = \text{diag}\left(\mathbf{b}^{(m)}_{k}\right)$, it is possible to define the cascaded channel $\mathbf{\tilde{h}}_{i,m,k}\in \mathbb{C}^{1 \times M}$ between the \ac{BS} and the $i$-th \ac{UE} through the $k$-th tile when employing the $m$-th configuration vector $\mathbf{b}^{(m)}_{k}$ as
\begin{equation}
\label{eq:risChan}
\mathbf{\tilde{h}}_{i,m,k} = \mathbf{t}_{i,k} \mathbf{B}^{(m)}_{k}\mathbf{S}_{k}.
\end{equation}
Accordingly, \eqref{eq:rx_irs3} can be written with a simplified notation as
\begin{equation}
{z}_{i,k,j} = \sum\limits_{m=1}^{N_u}\alpha_{m,k} \mathbf{\tilde{h}}_{i,m,k} \mathbf{x}_j.
\label{eq:rx_irs4}
\end{equation}
The $i$-th \ac{UE} is usually connected to the \ac{BS} also by the direct channel $\mathbf{\bar{h}}_{i} \in\mathbb{C}^{1 \times M}$, so the overall received signal relative to the $j$-th stream, {namely ${y}_{i,j}$, }includes ${w}_{i,j} = \mathbf{{\bar{h}}}_{i} \, \mathbf{x}_j$ as well, and can be formulated as
\begin{equation}
\begin{aligned}
{y}_{i,j} & = \sum\limits_{k=1}^{K}{z}_{i,k,j} + {w}_{i,j}= \left(\mathbf{{\bar{h}}}_{i} + \sum\limits_{k=1}^{K}\sum\limits_{m=1}^{N_u}\alpha_{m,k} \mathbf{\tilde{h}}_{i,m,k} \right) \mathbf{x}_j.
\end{aligned}
\label{eq:rx_total}
\end{equation} 
Let $\mathbf{\tilde{H}}_i \in \mathbb{C}^{N_u K \times M}$ be the matrix obtained by stacking the $N_uK$ cascaded channels $\mathbf{\tilde{h}}_{i,m,k}$ ($m = 1, \ldots, N_u$ and $k = 1, \ldots, K$) so that the $r$-th row of the matrix is expressed as $r = (k-1)N_u+m$. By denoting with $\boldsymbol{\alpha} =[\alpha_{1,1},\ldots,\alpha_{N_u, K}]\in \mathbb{C}^{1 \times N_uK}$ the row vector obtained by collecting all the coefficients $\alpha_{m,k}$ retaining the same ordering w.r.t. $k$ and $m$, {the received signal} ${y}_{i,j}$ can be written as
\begin{equation}
\begin{aligned}
{y}_{i,j} & = \left(\mathbf{{\bar{h}}}_{i} + \boldsymbol{\alpha} \mathbf{\tilde{H}}_i \right) \mathbf{x}_j .
\end{aligned}
\label{eq:rx_total2}
\end{equation}
{So t}he total channel $\mathbf{{h}}_{i} \left(\boldsymbol{\alpha}\right)\in \mathbb{C}^{1 \times M} $ between the BS and the $i$-th UE {
is given by} 
\begin{equation}\label{eq:totalChannel}
\mathbf{{h}}_{i} \left(\boldsymbol{\alpha}\right) = \mathbf{{\bar{h}}}_{i} + \boldsymbol{\alpha} \mathbf{\tilde{H}}_i, 
\end{equation}
 and the signal received at the $i$-th {UE} is the sum of the { signals} of all the users and the white Gaussian noise ${n}_i\in \mathcal{CN}\left(0,\sigma_i^2\right)$, i.e., 
\begin{equation}
\begin{aligned}
{y}_{i} =&\sum\limits_{j=1}^{N_u}{y}_{i,j} + {n}_i = \mathbf{{h}}_{i} (\boldsymbol{\alpha}) \mathbf{x}_i + \mathbf{{h}}_{i} (\boldsymbol{\alpha}) \sum\limits_{\substack{j=1,\\j \neq i}}^{N_u }\mathbf{x}_j + {n}_i. 
\end{aligned}
\label{eq:rx_total3}
\end{equation}
Accordingly, the \ac{SINR} for user $i$ is
\begin{equation}
\text{SINR}_i\left(\boldsymbol{\alpha},\mathbf{V}\right) = \frac{| \mathbf{h}_{i}\left(\boldsymbol{\alpha}\right) \mbf{v}_{i} | ^ 2}{\sum\limits_{j = 1, j \ne i }^{N_{u}} | \mathbf{h}_{i}\left(\boldsymbol{\alpha}\right) \mbf{v}_{j} | ^ 2 + \sigma_i^2 },
\end{equation}
where $\mathbf{V}=[\mbf{v}_{1},\ldots,\mbf{v}_{N_{u}}] \in \mathbb{C} ^ {M \times N_u}$ is the total precoding matrix of all the $N_{u}$ UEs{. Therefore, by modeling interference as noise, } the achievable rate for the $i$-th UE is
\begin{equation}
\label{eq:rate2.1}
R_{i}\left(\mathbf{V},\boldsymbol{\alpha}\right)= \log
\left(
{1} +\text{SINR}_i\left(\boldsymbol{\alpha},\mathbf{V}\right) 
\right).
\end{equation}
{ Before proceeding further, we recall the close relationship between the \ac{SINR} and the \ac{MMSE} that is exploited next. For the information symbol $\theta_i$ transmitted by the $i$-th \ac{UE} ($i=1,\ldots,N_u$), the \ac{MSE} ${E}_{i}\left(g_i;\boldsymbol{\alpha},\mathbf{V}\right)$ is defined as
\begin{equation}
\label{eq:MSEmat}
\begin{aligned}
{E}_{i}\left(g_i;\boldsymbol{\alpha},\mathbf{V}\right) & = \E\left\{ {\left( {{\txsymbol_i} - g^H_{i} {y}_{i}} \right){{\left( {{\txsymbol_i} - g^H_{i} {y}_{i}} \right)}^H}} \right\},
\end{aligned} 
\end{equation}
where $g^H_{i}$ is the receiver weight, so that the estimated received symbol for the \textit{i}-th \ac{UE} is $\hat{\theta}_i = g^H_{i}{y}_{i}$. Replacing \eqref{eq:rx_total3} in \eqref{eq:MSEmat} and exploiting the i.i.d. zero-mean assumption for the information symbols yields
\begin{equation}
\label{eq:MSEmat1}
\begin{aligned}
{E}_{i}\left(g_i;\boldsymbol{\alpha},\mathbf{V}\right)  = &~ 1+|g_i|^2\left(\sum\limits_{j = 1}^{N_{u}} | \mathbf{h}_{i}\left(\boldsymbol{\alpha}\right) \mbf{v}_{j}| ^ 2 + \sigma_i^2\right)\\&-2\Re\left[g_i^H \mathbf{h}_{i}\left(\boldsymbol{\alpha}\right) \mathbf{v}_i\right].
\end{aligned}
\end{equation}
Optimizing \eqref{eq:MSEmat1} w.r.t. $g_i$ yields the optimal receiver coefficient 
 \begin{align}
  \label{eq:optg}
g^*_i(\boldsymbol{\alpha},\mathbf{V}) = \frac{\mathbf{h}_{i}\left(\boldsymbol{\alpha}\right) \mbf{v}_{i}}{\sum\limits_{j = 1}^{N_{u}} | \mathbf{h}_{i}\left(\boldsymbol{\alpha}\right) \mbf{v}_{j}| ^ 2 + \sigma_i^2}, 
\end{align}
 so that, replacing \eqref{eq:optg} into \eqref {eq:MSEmat1}, yields the \ac{MMSE} 
\begin{align}
\label{eq:MMSE}
{E}^*_{i}\left(\boldsymbol{\alpha},\mathbf{V}\right) = & ~{E}_{i}\left(g^*_i\left(\boldsymbol{\alpha},\mathbf{V}\right);\boldsymbol{\alpha},\mathbf{V}\right) \\ = 
& ~1-\frac{| \mathbf{h}_{i}\left(\boldsymbol{\alpha}\right) \mbf{v}_{i}| ^ 2}{\sum\limits_{j = 1}^{N_{u}} | \mathbf{h}_{i}\left(\boldsymbol{\alpha}\right) \mbf{v}_{j} | ^ 2 + \sigma_i^2 } \triangleq \left(1+\text{SINR}_i\left(\boldsymbol{\alpha},\mathbf{V}\right)\right) ^{-1}. \nonumber
\end{align}}
\subsection{{
Basis Vector Design}} 
The basis vectors are designed such that $\mathbf{b}_k^{(m)}$ is the configuration of tile $k$ that maximizes the {amount of reflected power }towards node $m$ when the signal is transmitted by the BS and there is no direct path. This problem is extensively studied in the literature and many approaches have been proposed \cite{basar2019wireless, direnzo2020smart,Abrardo2021,abrardo2021MIMO, Wu2024Intelligent}. In our case, the solution is based on the assumption that the channel between the BS and the RIS is characterized by a strong \ac{LOS} component. This assumption is reasonable since the locations of the BS and the RIS are typically fixed and known, and they can be positioned appropriately to ensure a LOS connection that predominates over the multipath. In this setting, it is well known that an RIS configuration design that is reasonably close to optimal consists of directing the scattered signal toward the center of the BS antenna system, both in the near-field and far-field regimes \cite{bartoli2023spatial}. This design is also independent of the precoding vector utilized at the BS. To elaborate, let
 $\mathbf{K}_{k} \in \mathbb{C}^{P\times 1}$ be the vector encompassing the \ac{LOS} component of the link connecting the $k$-th tile of the \ac{RIS} to the center of the BS antenna. 
The basis functions are then chosen by solving $N_u$ sub-problems:
\begin{align} \label{eq:base_functions_set}
\mathbf{b}_k^{(m)} =\arg \max \limits_{\mathbf{b}}\left|\mathbf{t}_{m,k}\text{diag}\left(\mathbf{b}\right)\mathbf{K}_{k}\right|^2,
\end{align}
{for $m = 1,\ldots,N_u$. Assuming the constraint $\left|{b}^{(m)}_{k,p}\right| = 1$ for simplicity, the solution of \eqref{eq:base_functions_set} is}
\begin{equation}
{b}^{(m)}_{k,p} = e^{j\left(-\measuredangle{t_{m,k}(p)}-\measuredangle{K_{k,p}}\right)}.
\end{equation}
{Note that the phase shifts $\measuredangle{K_{k,p}}$, for $p = 1,\ldots,P$, are known and, as mentioned, correspond to the physican operation of focusing the signal emitted by the $k$-th tile of the RIS towards the center of the BS, and $\measuredangle{t_{m,k}(p)}$ denotes the phase shifts introduced by the link between the $k$-th tile and the $m$-th UE.}

\section{Problem Formulation}\label{sec:probelm_formulation}
One of the main advantages of employing RISs is that they can create an alternative path when there are obstacles on the direct link between the \ac{BS} and some \acp{UE}, so that users in poor coverage areas can achieve the required rates for communication with a much better power budget.  
In particular, since in the considered \ac{MIMO} scenario with a single multi-antenna transmitter and multiple single-antenna receivers, any rate constraint directly translates into a minimum \ac{SINR} constraint from \eqref{eq:rate2.1}, the optimization problem can be formulated as follows:
\begin{align}
\label{eq:minPw}    
& \min \limits_{\mathbf{V},\boldsymbol{\alpha}} \sum\limits_{i = 1}^{N_{u}} \lVert \mbf{v}_{i} \rVert ^2\\
\text{s.t.:} \nonumber \\  
&  \text{SINR}_i\left(\boldsymbol{\alpha},\mathbf{V}\right)  \geq \Sigma_i \quad 1 \le i\le N_u \tag{\ref{eq:minPw}.a} \label{eq:minPw_SINR} \\
& \quad \quad \left|\sum_{m=1}^{N_u}\alpha_{m,k} {b}_{k,p}^{(m)}\right|^2 = 1 
\tag{\ref{eq:minPw}.b} 
\label{eq:minPw_RIS} 
\end{align} 
where $\Sigma_i$ is the target \ac{SINR} corresponding to the target rate $R_i$ for the \textit{i}-th \ac{UE} and \eqref{eq:minPw_RIS} assumed no power losses. 

Problem \eqref{eq:minPw} is a power control problem and, by its nature, its solution depends on the condition of having a feasible set of solutions \cite{zander1992,yates1995}. Given our emphasis on algorithmic design, we will operate under the assumption that the problems we investigate are all feasible.
Due to \eqref{eq:minPw_SINR} {and \eqref{eq:minPw_RIS}}, problem \eqref{eq:minPw} is non-convex and, therefore, challenging to solve. Following the approach first proposed in \cite{wu2019intelligent}, we decompose \eqref{eq:minPw} into two sub-problems, both formulated with the goal of minimizing the expended power: \emph{precoding optimization} and \emph{tile configuration optimization}. 
Given that both optimization problems are oriented towards achieving the same objective, we can develop an iterative approach rooted in the \ac{AO} framework, which allows for the sequential refinement of solutions by optimizing one variable set at a time while holding the others fixed, thereby progressively converging towards a (local) optimal solution. Within this framework, iteration $t$ is composed by the sequential optimization of the precoding matrix and tile coefficients, whichare denoted as $\mathbf{V}^{(t)}=[\mbf{v}_{1}^{(t)},\ldots,\mbf{v}_{N_{u}}^{(t)}]$ and $\boldsymbol{\alpha}^{(t)}=[\alpha^{(t)}_{1,1},\ldots,\alpha^{(t)}_{N_u, K}]$, respectively.

\subsection{Precoding Optimization}
\label{Sec_beamf}
Given a fixed RIS configuration $\boldsymbol{\alpha}^{(t)}$, the optimal precoder $\mathbf{V}^{(t)}$ can be found by solving the power minimization problem \hyperlink{PO}{PO} in $\mathbf{V}$ \cite{wu2019intelligent}, as follows:
\begin{align}
  \label{eq:minPw_precoding}
  \hypertarget{PO}{}
\text{PO}: ~ &\mathbf{V}^{(t)}=\arg \min \limits_{\mathbf{V}} \sum\limits_{i = 1}^{N_{u}} \lVert \mbf{v}_{i} \rVert ^2\\
\text{s.t.:} \nonumber \\
&  \text{SINR}_i\left(\boldsymbol{\alpha}^{(t)},\mathbf{V}\right)  \geq \Sigma_i \quad 1\le i\le N_u. \tag{\ref{eq:minPw_precoding}.a} \label{eq:minPw_precoding_SINR}  
\end{align}
In this formulation, well-known for conventional power minimization problems in the multiuser \ac{MIMO} downlink broadcast channel, problem \hyperlink{PO}{PO} can be viewed as an instance of an \ac{SOCP} problem and, as such, is convex and its solution can be found by employing standard solvers.
By the nature of the problem and the properties of power control, we can see that for any feasible solution (and hence also for the optimal one) all the \ac{SINR} constraints are binding, as 
clarified in the following remark.

\textbf{Remark 1}.\hypertarget{Remark1}{} \emph{The solution of \hyperlink{PO}{PO} always satisfies the constraints in \eqref{eq:minPw_precoding_SINR} with equality}. The proof is reported in Appendix \hyperlink{A}{A}.

 \subsection{Tiles Configuration Optimization}
{Here we assume that the beamforming matrix $\mathbf{V}^{(t)}$ is fixed. Then, the \ac{AO} framework consists of solving an optimization problem designed to configure the coefficients $\boldsymbol{\alpha}^{(t+1)}$ in order to increase} 
{the \ac{SINR} of each user.\\ 
To elaborate, the SINR constraints can be easily translated in MMSE constraints from \eqref{eq:MMSE}. Furthermore, by exploiting the same relationship, a convenient form for maximizing the users' SINR is to minimize the MMSE. However, since the \ac{MMSE} ${E}^*_{i}\left(\boldsymbol{\alpha},\mathbf{V}^{(t)}\right)$ in \eqref{eq:MMSE} is not convex in $\boldsymbol{\alpha}$, we instead employ the convex MSE $ {E}_{i}(g_{i}^{(t)};\boldsymbol{\alpha},\mathbf{V}^{(t)})$ given in \eqref{eq:MSEmat1}.
Consequently the objective function becomes the sum of MSEs, 
$ \sum_{i = 1}^{N_{u}} {E}_{i}\left(g_{i}^{(t)};\boldsymbol{\alpha},\mathbf{V}^{(t)}\right),$
where $g_{i}^{(t)}=g_{i}^*\left(\boldsymbol{\alpha}^{(t)},\mathbf{V}^{(t-1)}\right)$ is the optimal receiver coefficient found by employing \eqref{eq:optg} at step $t$. {In other words, the AO framework is applied for optimizing $\boldsymbol{\alpha}$ and $g_{i}$ as well, since once the best $\boldsymbol{\alpha}$ is found at step $t+1$, the MSE is further minimized by employing \eqref{eq:optg} to find the optimal $g_{i}^{(t+1)}$ and consequently ${E}^*_{i}\left(\boldsymbol{\alpha}^{(t+1)},\mathbf{V}^{(t)}\right)$.}}\\
{When implementing the AO framework as mentioned, it is necessary to tackle the non-convex \ac{LC} constraint \eqref{eq:minPw_RIS}. To this end, we consider the less stringent global convex constraint that the tile, as a whole, scatters an amount of power that is less than or equal to the total impinging power, i.e.,
\begin{equation}
\left\lVert\ \sum\nolimits_{m=1}^{N_u} \alpha_{m,k} \mathbf{b}_k^{(m)} \right\rVert ^2 \le P. 
\label{cons_GC}
\end{equation}
The constraint in \eqref{cons_GC} is referred to as \ac{GC} and is a convex constraint \cite{fotock2024energy}. To find a solution that fulfills the original \ac{LC} constraint, we adopt the approach of projecting the obtained solution onto the unit circle \cite{Abrardo2021}. Specifically, denoting the solution found imposing the \ac{GC} constraint as $\mathbf{b}_k^{(GC)}$, we obtain the \ac{LC}-constrained solution $\mathbf{b}_k^{(LC)}$ as follows:
\begin{equation}\label{Projection}
\mathbf{b}_k^{(LC)}=e^{j\angle{\mathbf{b}_k^{(GC)}}}.
\end{equation}
The projection is performed at the end of each iteration in Algorithm \ref{alg:tiles_and_precoder_opt}. }
\\ 
Based on the proposed approach, 
the RISs coefficients at step $t+1$ can be found by solving problem 
 \hyperlink{TO1}{TO1} as follows:
\begin{align}
\label{eq:minMSE}
\hypertarget{TO1}{}
\text{TO1} : ~& \boldsymbol{\alpha}^{(t+1)} = \text{arg}\min \limits_{\boldsymbol{\alpha}} \sum\limits_{i = 1}^{N_{u}} {E}_{i}\left(g_{i}^{(t)};\boldsymbol{\alpha},\mathbf{V}^{(t)}\right)\\
\text{s.t.:} \nonumber\\
& {E}_{i}\left(g_{i}^{(t)};\boldsymbol{\alpha},\mathbf{V}^{(t)}\right) \leq {E}^*_{i}\left(\boldsymbol{\alpha}^{(t)},\mathbf{V}^{(t)}\right),\tag{\ref{eq:minMSE}.a} \label{eq:minMSE_SINR}\\
& \quad \quad \left\lVert\ \sum\nolimits_{m=1}^{N_u} \alpha_{m,k} \mathbf{b}_k^{(m)} \right\rVert ^2 \le P.
\tag{\ref{eq:minMSE}.b} 
\label{eq:minPw_RIS2}
\end{align}
Problem \hyperlink{TO1}{TO1} is convex in $\boldsymbol{\alpha}$  and can be solved by using standard convex solvers. However, in Section \ref{optSolution} we will discuss the low complexity algorithm for finding its optimal solution. It is important to underscore that minimizing the sum of the \acp{MSE}, which approximately translate into minimizing the sum of the reciprocal of the \acp{SINR} with individual constraints on each SINR, inherently promotes fairness, as it penalizes solutions that result in very low SINRs for certain users.
\begin{algorithm}
\small
    \SetAlgoLined
    \textbf{Initialization }\\
    -Set an initial random tile configuration $\boldsymbol{\alpha}^{(t)}$ for $t = 1$; \\
    -Set an arbitrarily small value $\varepsilon$, $\Delta=1$;\\ 
    \While{$\Delta > \varepsilon$, \text{given} $\boldsymbol{\alpha}^{(t)}$ }{    
               -  \textbf{Precoding Optimization}\\
               - Compute $\mbf{V}^{(t)}$ as a solution of \hyperlink{PO}{PO};\\
 - \textbf{Tiles Configuration Optimization}\\
       - Compute $g_{i}^{(t)}=g_{i}^*\left(\boldsymbol{\alpha}^{(t)},\mathbf{V}^{(t)}\right)$ as in \eqref{eq:optg};\\
        - Compute $\boldsymbol{\alpha}^{(t+1)}$ by solving problem \hyperlink{TO1}{TO1};\\
        - \textbf{$\Delta$ and $t$ Update }\\
         - $\Delta= \left \lvert \sum\limits_{i = 1}^{N_{u}}  \frac{\lVert\mbf{v}_i^{(t)} - \mbf{v}_i^{(t-1)}\rVert}{\lVert \mbf{v}_i^{(t-1)}\rVert}  \right \rvert;$\\
        - t = t + 1;\\
        }
\caption{Precoder and Tile Configuration Optimization}
\label{alg:tiles_and_precoder_opt}
\end{algorithm}
\subsection{Convergence Analysis}
{
The convergence of the proposed approach can be proved form the following chain of inequalities:
\begin{equation}
\begin{aligned}
\label{eq:minMSE2}
{E}^*_{i}\left(\boldsymbol{\alpha}^{(t+1)},\mathbf{V}^{(t)}\right)& \overset{\text{(a)}}{\le} 
{E}_{i}\left(g_{i}^{(t)};\boldsymbol{\alpha}^{(t+1)},\mathbf{V}^{(t)}\right)\\
&\overset{\text{(b)}}{\le}{E}^*_{i}\left(\boldsymbol{\alpha}^{(t)},\mathbf{V}^{(t)}\right)\overset{\text{(c)}}{=}\frac{1}{1+\Sigma_i}.
\end{aligned}
\end{equation}
Inequality (a) is a consequence of the fact that ${E}^*_{i}\left(\boldsymbol{\alpha}^{(t+1)},\mathbf{V}^{(t)}\right)$ is obtained by minimizing the MSE with the optimized $g_i^{(t+1)}$ via \eqref{eq:optg}; inequality (b) directly follows from the constraints in \ref{eq:minMSE_SINR}; finally (c) stems from Remark \hyperlink{Remark1}{1}. Thus, by exploiting \eqref{eq:MMSE} and \eqref{eq:minMSE2}, we obtain the SINR at iteration $t+1$:
\begin{equation}
\text{SINR}_i\left(\boldsymbol{\alpha}^{(t+1)},\mathbf{V}^{(t)}\right) = {E}^*_{i}\left(\boldsymbol{\alpha}^{(t+1)},\mathbf{V}^{(t)}\right)^{-1}-1\ge \Sigma_i.
\end{equation}
Since all the \acp{SINR} are equal to or greater than the designated target values, any new iteration of \hyperlink{PO}{PO} decreases the amount of consumed power. Since the power decreases at every iteration of the outer \ac{AO} algorithm, convergence is ensured to a local optimum of \eqref{eq:minPw}. 
The whole procedure is summarized in Algorithm \ref{alg:tiles_and_precoder_opt}.}

\section{A Low Complexity Solution for TO1} 
\label{optSolution}
{ 
In this section, we focus on a single instance of the problem and omit the iteration index $t$ for the sake of clarity. To obtain a low complexity solution to the problem \hyperlink{TO1}{TO1}, we start by reformulating the \ac{GC} constraints in \eqref{cons_GC}. To elaborate, for every element $p$ of tile $k$ we introduce the vector $\boldsymbol{\beta}_{k,p}=[b_{k,p}^{(1)},\dots, b_{k,p}^{(N_u)}]^T$, which collects the coefficients from the $N_u$ basis vectors $\mathbf{b}_k^{(m)}$ and define the vector ${\mathbf{q}}_{k,p}   = \mathbf{e}_k \otimes {\boldsymbol{\beta}}_{k,p}\in \mathbb{C}^{KN_u}$, where $\mathbf{e}_k $ is the $K$-dimensional unit vector with all zeros except in position $k$ where it is one. 
{Defining $\mathbf{Q}_{k,p} = \mathbf{q}_{k,p} \mathbf{q}_{k,p}^H$ as a positive semi-definite matrix by construction and $\mathbf{\tilde{Q}}_{k}=\sum_{p = 1}^{P} \mathbf{Q}_{k,p}$, the \ac{GC} constraint in \eqref{cons_GC} for tile $k$ can be reformulated as 
\begin{equation}
\label{eq:relaxedConstraint}
\boldsymbol{\alpha} \mathbf{\tilde{Q}}_{k} \boldsymbol{\alpha}^H \le P.
    \end{equation}
By defining $\mathcal{E}_i = 1/(1 + \Sigma_i)$, the RIS configuration problem \hyperlink{TO1}{TO1} with the \ac{GC} constraints can then be rewritten as 
\begin{align}
\hypertarget{TO2}{}
\label{eq:convMinMSE}
\text{TO2} : ~& \min \limits_{\boldsymbol{\alpha}} \sum\limits_{i = 1}^{N_{u}} {E}_{i}\left(g_i;\boldsymbol{\alpha},\mathbf{V}\right) & \quad\\
\text{s.t.:} \nonumber\\
& {E}_{i}\left(g_i;\boldsymbol{\alpha},\mathbf{V}\right) - \mathcal{E}_i\leq 0 \quad 1\le i\le N_u, \tag{\ref{eq:convMinMSE}.a} \label{eq:convMinMSE_SINR}\\
&\boldsymbol{\alpha} \mathbf{\tilde{Q}}_{k}\boldsymbol{\alpha}^H-P \le 0, \quad 1\le k\le K. \tag{\ref{eq:convMinMSE}.b} \label{eq:conv_RIS}
\end{align}
} Problem \hyperlink{TO2}{TO2} is a standard convex problem for which it is possible to derive the expression of the Lagrangian dual function and its gradient in closed form. As a result, the problem can be solved with low complexity in the dual domain using the \ac{GAM} to maximize the Lagrangian dual function. The derivation of the Lagrangian dual function and the solution of the \ac{LDP} for problem \hyperlink{TO2}{TO2} is provided for completeness in Appendix \hyperlink{AppendixB}{B}.\\ 
It is important to mention that the solution for \hyperlink{TO2}{TO2} can be derived in the dual domain only when strong duality conditions hold. Typically, this condition is verified through Slater’s qualification conditions \cite{bertsekas2016nonlinear}, which unfortunately do not apply here. In this scenario, leveraging some considerations about the wireless propagation channel, we can justify the strong duality conditions by exploiting the convexity of the problem and the regularity of the solution. Interested readers are referred to Appendix \hyperlink{AppendixC}{C} for further exploration of the topic, where detailed derivations and discussions are provided.}

\subsection{Computational Complexity}
\label{sec:computational_complexity}
\begin{figure}
    \centering
    \includegraphics[width=0.7\linewidth]{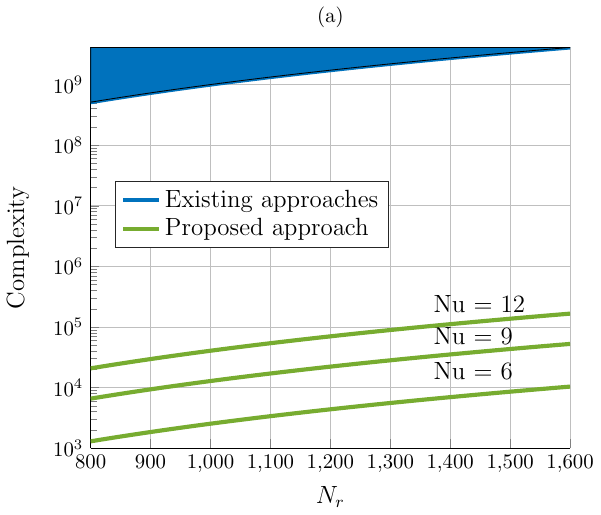}
    \includegraphics[width=0.7\linewidth]{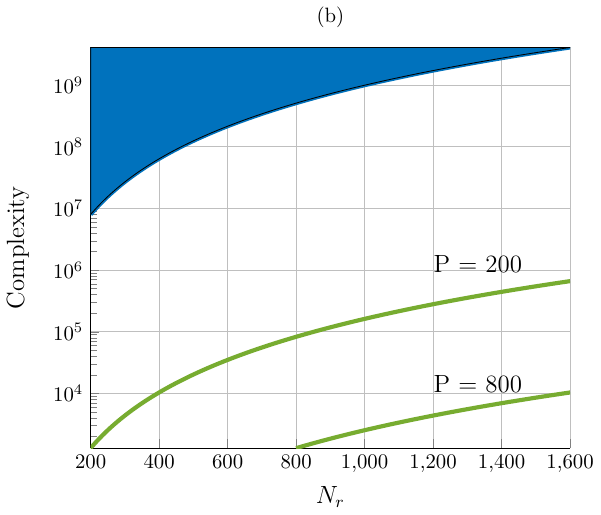}
    \caption{Computational complexity comparison for $P = 800$ and different values of $N_u$ (\ref{fig:ComputationalComplexity}a) and for $N_u = 6$ and different values of $P$ (\ref{fig:ComputationalComplexity}b). As for the existing approaches, the lower bound $N_r^3$ is considered. }
    \label{fig:ComputationalComplexity}
\end{figure}

{
The computational complexity of each iteration of the proposed algorithm is determined by the computational complexities of the problems \hyperlink{PO}{PO} and \hyperlink{TO2}{TO2}, i.e., the BS precoding and RIS configuration design. In the following, we focus on \hyperlink{TO2}{TO2}, traditionally the most computationally expensive due to the size of the problem \cite{wu2019intelligent}, 
and we compare its computational complexity against other state-of-the-art algorithms. 
All the examined methods reported next, utilize iterative optimization schemes and, to facilitate comparison, we present the complexity analysis for a single iteration of each algorithm. It is worth noting that, the number of iterations required for our proposed algorithm to converge is notably low, as will be illustrated in Section \ref{sec:results}.
As far as the benchmark schemes are concerned, we focus mainly on \cite{wu2019intelligent} and \cite{kumar2023novel}, as they are the most closely related methods as compared to ours. As shown in \cite{kumar2023novel}, the computational complexity of the algorithm in \cite{wu2019intelligent} is approximately $\mathcal{O}(N_r^7)$. More recent methods have been proposed to reduce this complexity, though it still remains of the order of $\mathcal{O}(N_r^\alpha)$, where $\alpha > 3$. 
}
\begin{table}
\small
\begin{center}
\makebox[\linewidth]{
\begin{tabular}{ |c|c|} 
 \hline
 \rowcolor{silver} Reference & Complexity\\
  \hline
     Proposed & $\mathcal{O}\left[\frac{N^3_r}{Z^3}\max\left(K ,N_u  \right)\right]$ \\ \hline
 \cite{wu2019intelligent} &  $\mathcal{O}(N_r^7)$\\ \hline
\cite{wang2022transmit} & $\mathcal{O}(N_r^4)$ \\ \hline
\cite{ma2022transmit} & $\mathcal{O}((N_u+1)N_r^{3.5})$ \\ \hline
\cite{zhoy2023transmit} &  $\mathcal{O}((N_u+1)N_r^{3.5})$\\ \hline
\cite{feng2023resource} &  $\mathcal{O}(N_r^{4.5})$\\ \hline
\cite{ren2023energy} & $\mathcal{O}(N_r^3)$ \\ \hline
\cite{kumar2023novel} & $\mathcal{O}(N_r^3)$ \\ \hline
 \end{tabular}
}
\end{center}
\caption{Computational complexity comparison for the existing and proposed approaches.}
\label{tab:Complexities}
\end{table}
{With respect to the proposed approach, the computational complexity of solving \hyperlink{TO2}{TO2} arises primarily from the calculation of the dual function gradients, $\nabla_{\boldsymbol{\lambda}}d(\boldsymbol{\lambda},\boldsymbol{\mu})$ and $\nabla_{\boldsymbol{\mu}}d(\boldsymbol{\lambda},\boldsymbol{\mu})$, shown in \eqref{eq:partDerg2} and \eqref{eq:partDerg5} of Appendix \hyperlink{AppendixB}{B}. We assume, as is customary, that both the computation of a matrix inverse and the multiplication of square matrices scale cubically with the number of elements, disregarding more efficient algorithms for matrix multiplication for simplicity. 
Under these assumptions, the complexity of computing the gradient $\nabla_{\boldsymbol{\lambda}}d(\boldsymbol{\lambda},\boldsymbol{\mu})$ can be expressed as $\mathcal{O}(N_u^4K^3)$, and for $\nabla_{\boldsymbol{\mu}}d(\boldsymbol{\lambda},\boldsymbol{\mu})$ as $\mathcal{O}(N_u^3K^4)$. Considering that the total number of scattering elements is $N_r = KP$, where $P$ is the number of RIS elements per tile and $K$ is the number of tiles, introducing the parameter $Z = \frac{P}{N_u}$ allows us to express the complexity of a single iteration of our proposed method as $\mathcal{O}\left[\frac{N_r^3}{Z^3} \max\left(K, N_u\right)\right]$. This formulation facilitates a fair comparison with existing methods.\\
Table \ref{tab:Complexities} presents a summary of the computational complexity of different methods available in the literature. Our proposed method demonstrates a complexity reduction of the order of $\frac{Z^3}{\max\left(K, N_u\right)}$ compared to existing solutions. This reduction becomes particularly significant in scenarios where the number of RIS elements per tile $P$ is large and the number of users, $N_u$, remains relatively small. For example, in the scenarios considered in Section \ref{sec:results}, $P = 800$ is set to 800. With $N_u = 10$, our proposed method achieves a complexity reduction by a factor of $\frac{80^3}{10} = 51200$ compared to the best algorithms available in the literature.\\
Figure \ref{fig:ComputationalComplexity} further illustrates the computational complexities of the proposed and existing algorithms a function of $N_u$ and $P$. For simplicity, we utilize $N_r^3$ as a lower bound for the computational complexities of the existing methods. This baseline is illustrated in blue in the figure. Despite this conservative baseline, the substantial reduction in computational complexity offered by our approach is apparent.} 
  
\section{Numerical Results}
\label{sec:results}

{
We consider a BS equipped with $M = 16$ antennas arranged in two linear arrays, each comprising 8 antennas.
The antenna elements of the BS and the unit cells of the RIS are spaced $\lambda/2$ apart.
The bandwidth of the considered signal is $30$ KHz and it is centered at $f_\mathrm{c}=28$ GHz, with a noise power spectrum density of $-174 ~\deci\bel\milli/\hertz$, a transmit power of $20~\deci\bel\milli$, a noise figure of $8~\deci\bel$, and the antenna gains in transmission and reception $\Gr=\Gt=3~\deci\bel$i. The results are obtained by averaging several instances for the spatial distribution of single-antenna UEs, which are placed at a height of 1 m, and are randomly deployed in a rectangular region of size $[0, 30] \times [0, 20]$ m.
As for the channels model, the direct BS-UE link $\mathbf{\bar{h}}_{i}$ is modelled according to the ABG \ac{NLOS} channel model introduced in \cite{5gchannelmodel2016}, since 
using RIS to support communications is recommended when the direct links are affected by strong attenuations, e.g., when the LOS component is obstructed by obstacles.
Conversely, as in \cite{palmucci2023two} and \cite{kumar2023novel}, the BS-RIS and RIS-UE connections are modeled as the combination of a \ac{LOS} Ricean component and a \ac{NLOS} Rayleigh multipath component. The Ricean factors of the BS-RIS and RIS-UE links, $\kappa_s$ and $\kappa_t$, are set to $50$, unless otherwise specified, modeling a strong LOS component in both links.}\\

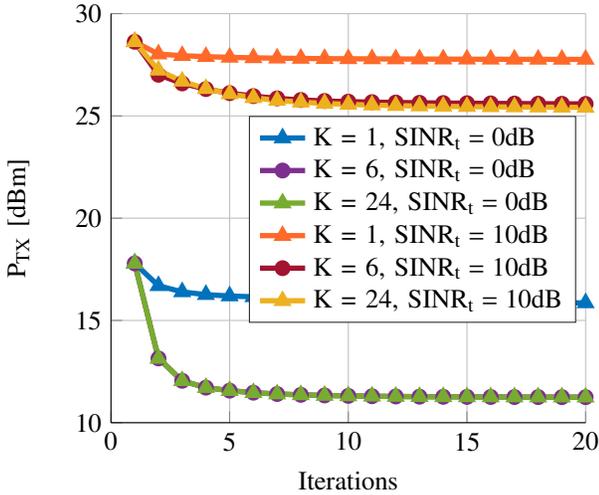
\begin{figure}
    \centering
%
%
\definecolor{mycolor1}{rgb}{0.49412,0.18431,0.55686}%
\definecolor{mycolor2}{rgb}{0.00000,0.44706,0.74118}%
\definecolor{mycolor3}{rgb}{0.46667,0.67451,0.18824}%
\definecolor{mycolor4}{rgb}{0.92941,0.69412,0.12549}%
\definecolor{mycolor5}{rgb}{1.00000,0.41176,0.16078}%
\definecolor{mycolor6}{rgb}{0.63529,0.07843,0.18431}%
\begin{tikzpicture}

\begin{axis}[%
width=0.7\linewidth,
scale only axis,
xmin=0,
xmax=20,
ymin=10,
ymax=30,
axis background/.style={fill=white},
axis x line*=bottom,
axis y line*=left,
xmajorgrids,
ymajorgrids,
title style={font=\bfseries},
xlabel= {Iterations},
ylabel = {$\text{P}_{\text{TX}}\text{ [dBm]}$},
legend style={legend cell align=left, align=left, draw=white!15!black, at ={(0.98,0.75)}}
]
\addplot [color=mycolor2, line width=2.0pt, mark=triangle*, mark options={solid}]
  table[row sep=crcr]{%
1	17.7890916878405\\
2	16.6962140979253\\
3	16.3959886411654\\
4	16.2613419606056\\
5	16.1943522384004\\
6	16.1551785951594\\
7	16.1219554170766\\
8	16.0903204627532\\
9	16.0616468189426\\
10	16.0359799511041\\
11	16.0017632377859\\
12	15.9609281396169\\
13	15.9395497194143\\
14	15.9213629155434\\
15	15.9060502130272\\
16	15.8922906159851\\
17	15.8819329410136\\
18	15.8731086620523\\
19	15.8649023162312\\
20	15.8569821264353\\
};
\addlegendentry{K = 1, $\text{SINR}_\text{t} = 0 \text{dB}$}

\addplot [color=mycolor1, line width=2.0pt,mark=*, mark options={solid}]
  table[row sep=crcr]{%
1	17.7890916877458\\
2	13.1437536747965\\
3	12.0485238074814\\
4	11.714879823503\\
5	11.5677582993298\\
6	11.4680318820507\\
7	11.4033244150163\\
8	11.362154728889\\
9	11.3329280703605\\
10	11.3112233417909\\
11	11.2948489565063\\
12	11.2823390098563\\
13	11.2730159237686\\
14	11.2665161911627\\
15	11.261766584374\\
16	11.2582043640664\\
17	11.2552134874184\\
18	11.252655536416\\
19	11.2504196166748\\
20	11.2484382055698\\
};
\addlegendentry{K = 6, $\text{SINR}_\text{t} = 0 \text{dB}$}

\addplot [color=mycolor3, line width=2.0pt, mark=triangle*, mark options={solid}]
  table[row sep=crcr]{%
1	17.7890916878405\\
2	13.1407903499158\\
3	12.0477830891688\\
4	11.7143520230153\\
5	11.5661888008844\\
6	11.4654987503526\\
7	11.4011009233092\\
8	11.3598834737931\\
9	11.3306486172797\\
10	11.309172439862\\
11	11.2929488942899\\
12	11.280719195222\\
13	11.2717236086647\\
14	11.2652380252011\\
15	11.2605722941213\\
16	11.256811364023\\
17	11.2537326934668\\
18	11.2510944617473\\
19	11.2487572407994\\
20	11.2464696973789\\
};
\addlegendentry{K = 24, $\text{SINR}_\text{t} = 0 \text{dB}$}

\addplot [color=mycolor5, line width=2.0pt, mark=triangle*, mark options={solid}]
  table[row sep=crcr]{%
1	28.6230549552925\\
2	28.022212918185\\
3	27.9451180332835\\
4	27.8967761855398\\
5	27.8633062373852\\
6	27.8397194272504\\
7	27.8227771142609\\
8	27.8093465339473\\
9	27.7996963704215\\
10	27.7931818477271\\
11	27.787379005104\\
12	27.7832168205371\\
13	27.7794336486262\\
14	27.7761452648331\\
15	27.7730253881795\\
16	27.7702740005325\\
17	27.7680092396444\\
18	27.7660013930496\\
19	27.7641224280785\\
20	27.7625741359738\\
};
\addlegendentry{K = 1, $\text{SINR}_\text{t} = 10 \text{dB}$}

\addplot [color=mycolor6, line width=2.0pt,mark=*, mark options={solid}]
  table[row sep=crcr]{%
1	28.6230549552925\\
2	27.0120417583056\\
3	26.5890040773458\\
4	26.3116093890291\\
5	26.1143496486676\\
6	25.9678445171715\\
7	25.8584209638214\\
8	25.7798532376313\\
9	25.7305689929545\\
10	25.6972877298332\\
11	25.6737014902549\\
12	25.6554451838036\\
13	25.6407035078152\\
14	25.6293213101705\\
15	25.6201741391128\\
16	25.6125473432611\\
17	25.6050774058949\\
18	25.5980234106755\\
19	25.5918978465263\\
20	25.5872079549051\\
};
\addlegendentry{K = 6, $\text{SINR}_\text{t} = 10 \text{dB}$}

\addplot [color=mycolor4, line width=2.0pt, mark=triangle*, mark options={solid}]
  table[row sep=crcr]{%
1	28.6230549516145\\
2	27.230542560255\\
3	26.6923950740448\\
4	26.3310523127322\\
5	26.0778845783297\\
6	25.8882791463929\\
7	25.7569393851204\\
8	25.6703729516924\\
9	25.6098033638656\\
10	25.5643680485387\\
11	25.5288995181416\\
12	25.5022971957985\\
13	25.4820655811323\\
14	25.4666632654405\\
15	25.4546458624874\\
16	25.444601947238\\
17	25.4359087623327\\
18	25.4282960264104\\
19	25.4212143979261\\
20	25.4145018609127\\
};
\addlegendentry{K = 24, $\text{SINR}_\text{t} = 10 \text{dB}$}

\end{axis}

\end{tikzpicture}%
    \caption{\textit{FF} case: system performance for different values of $K$, the SINR target is $0$ and $10 ~\deci\bel$, $N_u = 6.$ }
    \label{fig:changingK}
\end{figure}

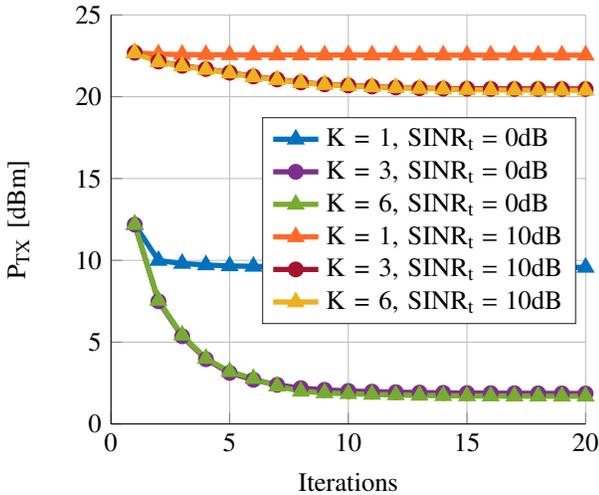
\begin{figure}
    \centering
%
%
\definecolor{mycolor1}{rgb}{0.49412,0.18431,0.55686}%
\definecolor{mycolor2}{rgb}{0.00000,0.44706,0.74118}%
\definecolor{mycolor3}{rgb}{0.46667,0.67451,0.18824}%
\definecolor{mycolor4}{rgb}{0.92941,0.69412,0.12549}%
\definecolor{mycolor5}{rgb}{1.00000,0.41176,0.16078}%
\definecolor{mycolor6}{rgb}{0.63529,0.07843,0.18431}%
\begin{tikzpicture}

\begin{axis}[%
width=0.7\linewidth,
scale only axis,
xmin=0,
xmax=20,
ymin=0,
ymax=25,
axis background/.style={fill=white},
axis x line*=bottom,
axis y line*=left,
xmajorgrids,
ymajorgrids,
title style={font=\bfseries},
xlabel= {Iterations},
ylabel = {$\text{P}_{\text{TX}}\text{ [dBm]}$},
legend style={legend cell align=left, align=left, draw=white!15!black, 
at = {(0.98,0.75)}}
]
\addplot [color=mycolor2,line width=2.0pt, mark=triangle*, mark options={solid}]
  table[row sep=crcr]{%
1	12.1824936491854\\
2	9.98847193058049\\
3	9.80639522240589\\
4	9.71694314071674\\
5	9.65816162400695\\
6	9.62904726942025\\
7	9.61226796591638\\
8	9.60306061305415\\
9	9.59512329962982\\
10	9.59005799411153\\
11	9.58612662128271\\
12	9.58290521788329\\
13	9.58001134492397\\
14	9.57775668749322\\
15	9.57580482462473\\
16	9.57402464112109\\
17	9.57234653033014\\
18	9.57074499654485\\
19	9.56920809333663\\
20	9.56793839222292\\
};
\addlegendentry{K = 1, $\text{SINR}_\text{t} = 0 \text{dB}$}

\addplot [color=mycolor1,line width=2.0pt, mark=*, mark options={solid}]
  table[row sep=crcr]{%
1	12.1824936456688\\
2	7.49592424590821\\
3	5.36116800371465\\
4	3.95536103898785\\
5	3.11877333498778\\
6	2.69790101859261\\
7	2.3879736944709\\
8	2.19002154682841\\
9	2.08554945691039\\
10	2.01767795229278\\
11	1.97468574873596\\
12	1.94237802789141\\
13	1.91892270589236\\
14	1.90174847928398\\
15	1.8898505690706\\
16	1.88192075765622\\
17	1.87572422195191\\
18	1.87038045531409\\
19	1.86578063694881\\
20	1.86164810802808\\
};
\addlegendentry{K = 3, $\text{SINR}_\text{t} = 0 \text{dB}$}

\addplot [color=mycolor3,line width=2.0pt, mark=triangle*, mark options={solid}]
  table[row sep=crcr]{%
1	12.1824936455565\\
2	7.54273448246975\\
3	5.4183210212166\\
4	4.00350764906662\\
5	3.19587552349152\\
6	2.77233716288647\\
7	2.30775147739395\\
8	1.99497883758225\\
9	1.8946708578666\\
10	1.84354817187479\\
11	1.80418613178555\\
12	1.772507357962\\
13	1.74538162009317\\
14	1.72366579001271\\
15	1.71192882369559\\
16	1.70569383019334\\
17	1.70121379960457\\
18	1.69794098703098\\
19	1.69505366866258\\
20	1.69264025455508\\
};
\addlegendentry{K = 6, $\text{SINR}_\text{t} = 0 \text{dB}$}

\addplot [color=mycolor5, line width=2.0pt, mark=triangle*, mark options={solid}]
  table[row sep=crcr]{%
1	22.67806172679\\
2	22.6017178696235\\
3	22.5762837844496\\
4	22.5650217080625\\
5	22.5579509224655\\
6	22.5532611468189\\
7	22.5500592190856\\
8	22.5479819258559\\
9	22.5464523862084\\
10	22.5451275242033\\
11	22.543932571685\\
12	22.5428635307355\\
13	22.5418961604563\\
14	22.5409957932765\\
15	22.5401499091126\\
16	22.5393542881032\\
17	22.5386025516133\\
18	22.5378854846136\\
19	22.5372115963278\\
20	22.5365861056261\\
};
\addlegendentry{K = 1, $\text{SINR}_\text{t} = 10 \text{dB}$}

\addplot [color=mycolor6, line width=2.0pt, mark=*, mark options={solid}]
  table[row sep=crcr]{%
1	22.6880617284884\\
2	22.1502772063684\\
3	21.8900951658406\\
4	21.6926897253408\\
5	21.4663763101286\\
6	21.2390469921458\\
7	21.0547512313299\\
8	20.8743270804844\\
9	20.7550852142373\\
10	20.6893896955005\\
11	20.622792481485\\
12	20.5615691796341\\
13	20.5193280423424\\
14	20.4936974771339\\
15	20.4907355196799\\
16	20.4777680782562\\
17	20.4755614133056\\
18	20.4736630519641\\
19	20.4719294016255\\
20	20.4702483386035\\
};
\addlegendentry{K = 3, $\text{SINR}_\text{t} = 10 \text{dB}$}

\addplot [color=mycolor4, line width=2.0pt, mark=triangle*, mark options={solid}]
  table[row sep=crcr]{%
1	22.6780617284884\\
2	22.1302772063684\\
3	21.8700951658406\\
4	21.6626897253408\\
5	21.4463763101286\\
6	21.2090469921458\\
7	21.0347512313299\\
8	20.8543270804844\\
9	20.7350852142373\\
10	20.6693896955005\\
11	20.602792481485\\
12	20.5515691796341\\
13	20.4993280423424\\
14	20.4636974771339\\
15	20.4353825465285\\
16	20.4152339096532\\
17	20.4026455203867\\
18	20.3957892225713\\
19	20.3931074911874\\
20	20.3913703843833\\
};
\addlegendentry{K = 6, $\text{SINR}_\text{t} = 10 \text{dB}$}

\end{axis}
\end{tikzpicture}%
    \caption{\textit{NF} case: system performance for different values of $K$, the SINR target is $0$ and $10 ~\deci\bel$, $N_u = 3.$ }
    \label{fig:changingKsingleTile}
\end{figure}

{The first comparison we present aims to evaluate different choices regarding the number of tiles per RIS for various positions of the RISs and the BS, as well as for different numbers of nodes.} In this regard, we note that, as shown in Section \ref{sec:computational_complexity}, the complexity of the proposed approach depends, for a given number of scattering elements $N_r$, on the number of tiles $K$. Specifically, the complexity increases with $K^3$, and thus reducing $K$ provides a significant complexity advantage. On the other hand, a smaller value of $K$ also impacts the performance, as it reduces the number of basis vectors, i.e., the size of the search space. Therefore, in order to find a good trade-off between complexity and performance, we consider two different simulation scenarios in the near and far field regimes, which are referred to as \textit{NF} and \textit{FF}, respectively. In both scenarios, we consider a total number of scattering elements $N_r = 4800$, but with different configurations and numbers of RISs. In particular, in the \textit{NF} scenario, the scattering elements are arranged in a single $20 \times 240$ RIS, i.e., $Q = 1$, whose center is placed at the position $(15, 0, 3)$, and the RIS is aligned along the $(x, z)$ plane. As for the BS, it is positioned very close to the RIS, at $(16, 4, 2)$, with the antennas aligned in the $(y, z)$ plane. In this way, communication occurs within the near-field region of both the BS-RIS and RIS-UE links.
In the \textit{FF} scenario, the BS is positioned at $(30, 15, 2)$, 
and $Q = 6$ RISs are considered, each composed of $40 \times 20 = 800$ scattering elements. Two RISs are placed at the positions $(0, 20, 3)$ and $(0, 10, 3)$ in the $(y, z)$ plane, while the others are located in the $(x, z)$ plane at the positions $(3, 0, 3)$, $(13, 0, 3)$, $(8, 30, 3)$, and $(18, 30, 3)$.
The NLOS model is based on the \ac{IO} channel for the \textit{NF} scenario, while for
the \textit{FF} case, the \ac{IO} and \ac{SM} channels are considered with probabilities of 70\% and 30\%, respectively. It is worth noting that, although both cases refer to an NLOS situation, the \ac{IO} scenario is characterized by a stronger attenuation with distance, while the \ac{SM} scenario is characterized by a qualitatively better direct connection. Furthermore, the results are obtained by considering the GC constraint in \eqref{cons_GC}, with subsequent projection of the solution onto the unit circle according to \eqref{Projection}. 
The results are obtained through Monte Carlo simulations for $N_u = 3$ and $N_u = 6$ UEs for the \textit{NF} and \textit{FF} cases, respectively, where each instance considers different random UE positions and varying NLOS components. The overall performance is reported in terms of the required average transmit power expressed in $\deci \bel \milli$, where the average is computed over all the simulated instances.
The results are reported in Figures \ref{fig:changingK} and \ref{fig:changingKsingleTile} for the \textit{FF} and \textit{NF} cases, respectively. We compare the transmit power $P_{\text{TX}}$ versus the number of iterations of the algorithm for different values of $K$, considering target SINR values of $0~\deci\bel$ and $10~\deci\bel$.
{
Specifically, in the \textit{FF} scenario, we test the performance for  $K = [1, 6, 24]$ while for the \textit{NF} scenario, we consider $K = [1, 3, 6]$. The results show that, for the chosen values of 
$N_u$, specifically $N_u = 3$ and $N_u = 6$ for the NF and FF cases, respectively, the minimum values of $K$ that allow for achieving maximum performance are, respectively, $K = 3$
and $K = 6$, meaning $K = N_u$ seems to be a good trade-off between performance and complexity in these cases. It is important to note that the optimal selection of the number of tiles $K$ will generally depend on several factors, such as the specific scenario, the communication channel characteristics, and the number of nodes involved. On the other hand, determining the optimal value of $K$ is beyond the scope of this work. In the subsequent figures, unless otherwise indicated, we focus on the results derived in the aforementioned \textit{FF} scenario.}
 \begin{figure}
    \centering
%
%
\definecolor{mycolor2}{rgb}{0.00000,0.44700,0.74100}%
\definecolor{mycolor1}{rgb}{0.46600,0.67400,0.18800}%
\definecolor{mycolor2}{rgb}{0.92900,0.69400,0.12500}
\begin{tikzpicture}

\begin{axis}[%
width=0.6\linewidth,
height = 0.4\linewidth,
scale only axis,
xmin=0,
xmax=20,
xlabel style={font=\color{white!15!black}},
xlabel={Iterations},
ymin=11,
ymax=18,
ylabel style={font=\color{white!15!black}},
ylabel={$\text{P}_{\text{TX}}\text{ [dBm]}$},
axis background/.style={fill=white},
title style={font=\bfseries},
title={SINR target 0 dB},
axis x line*=bottom,
axis y line*=left,
xmajorgrids,
ymajorgrids,
legend style={legend cell align=left, align=left, draw=white!15!black}
]
\addplot [color=mycolor1, line width=2.0pt,mark=*, mark options={solid}]
  table[row sep=crcr]{%
1	17.1417665288723\\
2	12.7752387814318\\
3	11.8671582933802\\
4	11.6351451226059\\
5	11.5372416403603\\
6	11.481903071619\\
7	11.4442005824623\\
8	11.4149907319237\\
9	11.3936979763527\\
10	11.3765848254162\\
11	11.3615117317798\\
12	11.3474135663442\\
13	11.3361543659187\\
14	11.3281022241288\\
15	11.3220610603997\\
16	11.3172805555589\\
17	11.3132458988437\\
18	11.309631390376\\
19	11.3063150249788\\
20	11.3030680849682\\
};
\addlegendentry{Unit circle projection}

\addplot [color=mycolor2, line width=2.0pt,mark=*, mark options={solid}]
  table[row sep=crcr]{%
1	17.1417665288723\\
2	12.435124298172\\
3	11.5989839875784\\
4	11.3841052875944\\
5	11.2944444963769\\
6	11.2477740526771\\
7	11.2190146213526\\
8	11.1989965119278\\
9	11.1839671183606\\
10	11.1719090598224\\
11	11.1618917191571\\
12	11.1533671660971\\
13	11.1460350422393\\
14	11.1396821796923\\
15	11.1340258882408\\
16	11.128675090409\\
17	11.1235524684417\\
18	11.1186114047119\\
19	11.1141813746218\\
20	11.110515012704\\
};
\addlegendentry{No Projection}

\end{axis}

\end{tikzpicture}%
    \caption{Comparison for the case studies with and without projection on the unitary circle, assuming the SINR target of 0 $\deci\bel$ and $K = Q = 6$.}
    \label{fig:ConSenzaProj}
\end{figure}
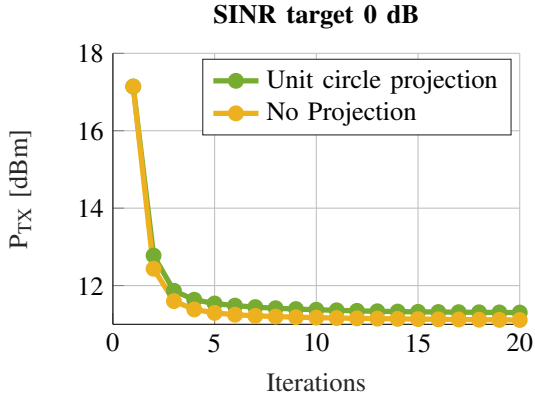

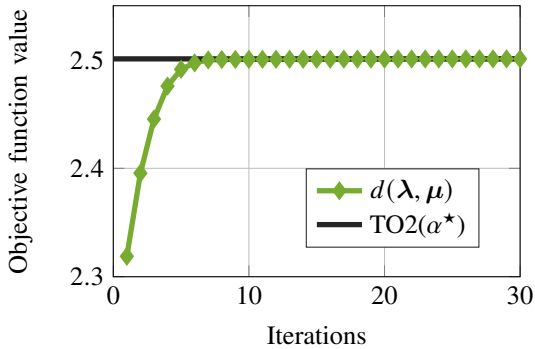
\begin{figure}
    \centering
%
%
\definecolor{mycolor3}{rgb}{0.00000,0.44700,0.74100}%
\definecolor{mycolor1}{rgb}
{0.46600,0.67400,0.18800}%
\definecolor{mycolor2}{rgb}{0.92900,0.69400,0.12500}%
\definecolor{mycolor4}{rgb}{0.49400,0.18400,0.55600}%
\definecolor{mycolor4}{rgb}{0.85000,0.32500,0.09800}
\begin{tikzpicture}

\begin{axis}[%
width=0.6\linewidth,
height = 0.4\linewidth,
scale only axis,
xmin=0,
xmax=30,
ymin=2.3,
ymax=2.55,
ylabel= {Objective function value},
xlabel = {Iterations},
xmajorgrids,
ymajorgrids,
axis background/.style={fill=white}, 
legend style={legend cell align=left, align=left, draw=white!15!black, at = {(0.9, 0.4)}}
]
\addplot [color=mycolor1, line width=2.0pt, mark=diamond*, mark options={solid}]
  table[row sep=crcr]{%
1	2.31885031417592\\
2	2.3955208697635\\
3	2.4454182086633\\
4	2.47586667119792\\
5	2.49132384309154\\
6	2.49727583004573\\
7	2.49995914823146\\
8	2.500348873828\\
9	2.50045198852344\\
10	2.50049815086639\\
11	2.50053117525942\\
12	2.50056012874322\\
13	2.50058749557407\\
14	2.50061417342611\\
15	2.50064053933458\\
16	2.50066676158793\\
17	2.50069291711532\\
18	2.50071904147466\\
19	2.50074515121083\\
20	2.50077125406004\\
21	2.50079735365462\\
22	2.50082345170627\\
23	2.50084954902431\\
24	2.50087564599259\\
25	2.50090174279369\\
26	2.50092783951467\\
27	2.50095099710008\\
28	2.50096559686048\\
29	2.50097068974812\\
30	2.50097068975188\\
31	2.50097068975294\\
};
\addlegendentry{$d(\boldsymbol{\lambda},\boldsymbol{\mu})$}
\addplot [color=white!15!black, line width=2.0pt ]
  table[row sep=crcr]{%
0	2.50097068975294\\
35	2.50097068975294\\
};
\addlegendentry{\hyperlink{TO2}{TO2}($\alpha^\star$)}
\end{axis}

\end{tikzpicture}%
    \caption{Convergence of the Lagrangian dual function}
    \label{fig:ConvergenceLagrangianDual}
\end{figure}

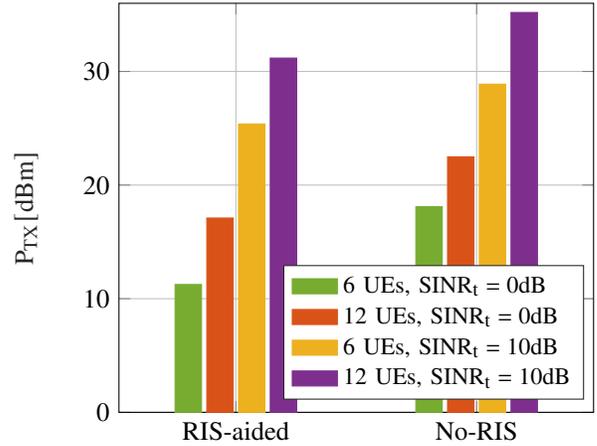
\begin{figure}
    \centering
%
%
\definecolor{mycolor2}{rgb}{0.00000,0.44700,0.74100}%
\definecolor{mycolor1}{rgb}{0.46600,0.67400,0.18800}%
\definecolor{mycolor2}{rgb}{0.92900,0.69400,0.12500}
\definecolor{mycolor4}{rgb}{0.49400,0.18400,0.55600}%
\definecolor{mycolor3}{rgb}{0.85000,0.32500,0.09800}
\begin{tikzpicture}

\begin{axis}[%
width=0.7\linewidth,
scale only axis,
bar shift auto,
xmin=0.514285714285714,
xmax=2.48571428571429,
xtick=data,
xticklabels={RIS-aided, No-RIS},
ymin=0,
ymax=36,
xmajorgrids,
ymajorgrids,
ylabel style={font=\color{white!15!black}},
ylabel={$\text{P}_{\text{TX}} [\text{dBm}]$},
title style={font=\bfseries},
axis background/.style={fill=white},
legend style={legend cell align=left, align=left, draw=white!15!black, at = {(0.98,0.36)}, font = \small}
]
\addplot[ybar,  bar width=10, draw=mycolor1,fill=mycolor1, area legend] table[row sep=crcr] {%
1 11.249953907355\\
2 18.0952245373446\\
};
\addlegendentry{6 UEs, $\text{SINR}_\text{t} = 0 \text{dB}$}

\addplot[ybar,  bar width=10, draw=mycolor3,fill=mycolor3, area legend] table[row sep=crcr] {%
1 17.0996871204392\\
2 22.48704307759890\\
};
\addlegendentry{12 UEs, $\text{SINR}_\text{t} = 0 \text{dB}$}
\addplot[ybar,  bar width=10, draw=mycolor2,fill=mycolor2,area legend] table[row sep=crcr] {%
1 25.37575035281007\\
2 28.88592564073070\\
};
\addlegendentry{6 UEs, $\text{SINR}_\text{t} = 10 \text{dB}$}
\addplot[ybar,  bar width=10, draw=mycolor4,fill=mycolor4,area legend] table[row sep=crcr]{%
1 31.177416444362071\\
2 35.193881524770011\\
};
\addlegendentry{12 UEs, $\text{SINR}_\text{t} = 10 \text{dB}$}

\end{axis}

\end{tikzpicture}%
    \caption{Required transmit power to achieve the target SINRs of $0$ and $10 ~\deci\bel $ for the RIS-aided and No-RIS cases with 6 and 12 UEs.}
    \label{fig:multiUE}
\end{figure}

In  Figure \ref{fig:ConSenzaProj}, we report the average transmitted power $P_{\text{TX}}$ for a target SINR of $0~\deci\bel$ and {for $N_u = 6$}, as a function of the number of iterations of the AO algorithm. From the figure, it can be observed that the performance degradation induced by the projection on the unit circle is minimal. Accordingly, in the following figures, the simulation results are obtained for the GC-type constraint with subsequent projection onto the unit circle. 
In Figure \ref{fig:ConvergenceLagrangianDual} we study the claims made in Section \ref{optSolution}. More specifically, the figure shows the Lagrangian dual function $d(\boldsymbol{\lambda},\boldsymbol{\mu})$ as a function of the number of iterations for a given instance of the problem. As shown in the figure, 
already after $n = 8$ iterations the duality gap approaches zero and hence convergence to the optimal value of the primal \hyperlink{TO2}{TO2} problem is guaranteed. Similar trends are observed in all the examined cases in the following. 
To demonstrate the effectiveness of deploying RISs in the considered scenario, Figure \ref{fig:multiUE} compares the case studied in which RISs are present or absent. Specifically, the results are obtained for $N_u = [6,12]$ and they illustrate the transmitted power required to achieve the target \ac{SINR}.
The results demonstrate that the deployment of optimized RISs in the system significantly reduces the power required to achieve the target \ac{SINR}. This highlights the importance of integrating RISs into power-efficinet wireless networks. As expected, increasing the number of \acp{UE} necessitates more power to maintain the same \ac{SINR} target.

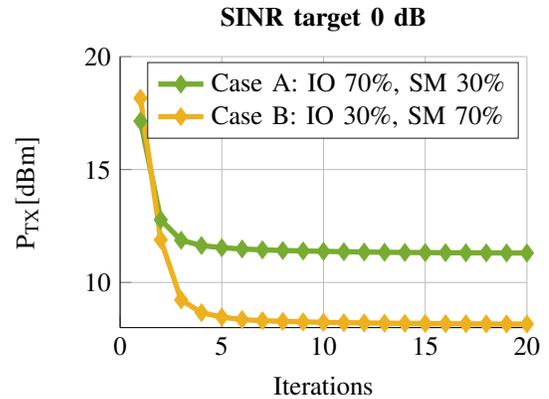
\begin{figure}
    \centering
%
\definecolor{mycolor2}{rgb}{0.00000,0.44700,0.74100}%
\definecolor{mycolor1}{rgb}{0.46600,0.67400,0.18800}%
\definecolor{mycolor3}{rgb}{0.85098,0.32549,0.09804}%
\definecolor{mycolor2}{rgb}{0.92900,0.69400,0.12500}
\begin{tikzpicture}

\begin{axis}[%
width=0.6\linewidth,
height = 0.4\linewidth,
scale only axis,
xmin=0,
xmax=20,
ymin=8,
ymax=20,
axis background/.style={fill=white},
axis x line*=bottom,
axis y line*=left,
xmajorgrids,
ymajorgrids,
title style={font=\bfseries},
title = {SINR target 0 dB},
xlabel= {Iterations},
ylabel={$\text{P}_{\text{TX}} [\text{dBm}]$},
legend style={legend cell align=left, align=left, draw=white!15!black}
]
\addplot[color=mycolor1, line width=2.0pt, mark=diamond*, mark options={solid}]
  table[row sep=crcr]{%
1	17.1417665288723\\
2	12.7752387814318\\
3	11.8671582933802\\
4	11.6351451226059\\
5	11.5372416403603\\
6	11.481903071619\\
7	11.4442005824623\\
8	11.4149907319237\\
9	11.3936979763527\\
10	11.3765848254162\\
11	11.3615117317798\\
12	11.3474135663442\\
13	11.3361543659187\\
14	11.3281022241288\\
15	11.3220610603997\\
16	11.3172805555589\\
17	11.3132458988437\\
18	11.309631390376\\
19	11.3063150249788\\
20	11.3030680849682\\
};
\addlegendentry{Case A: IO 70\%, SM 30\%}

\addplot [color=mycolor2, line width=2.0pt, mark=diamond*, mark options={solid}]
  table[row sep=crcr]{%
1	18.1576006084198\\
2	11.8861270391545\\
3	9.22784744956089\\
4	8.66372367384303\\
5	8.46599724128139\\
6	8.35698878245886\\
7	8.30476573571827\\
8	8.27651965244539\\
9	8.2534453117533\\
10	8.23396666043234\\
11	8.2185965214178\\
12	8.2063059216601\\
13	8.19626662443231\\
14	8.18807895399502\\
15	8.18079750365551\\
16	8.17456784987596\\
17	8.16897830991703\\
18	8.16427815718265\\
19	8.16030541201706\\
20	8.15695384511353\\
};
\addlegendentry{Case B: IO 30\%, SM 70\%}

\end{axis}

\end{tikzpicture}%
    \caption{Comparison in terms of BS-US link attenuation for the SINR target of $0 ~ \deci\bel$.}
    \label{fig:IOSMInvertiti}
\end{figure} 

A further comparison is presented in Figure \ref{fig:IOSMInvertiti}, where $N_u = 6$, and the direct BS-UE link is modeled according to two different cases. In Case A, considered in previous simulations, the IO and SM models were assumed with probabilities of $70\%$ and $30\%$,
respectively. In Case B, the probabilities are inverted. As previously mentioned, the IO model is characterized by a  stronger attenuation with the transmission distance. Thus, when it occurs with higher probability (Case A), worse results in terms of power minimization are expected, providing evidence of the role played by the direct BS-UE link for the considered scenario. 

In the next, we focus our attention on two benchmark schemes that are the most closely related to this research work.
\begin{figure}
    \centering
%
%
\definecolor{mycolor2}{rgb}{0.00000,0.44700,0.74100}%
\definecolor{mycolor1}{rgb}{0.46600,0.67400,0.18800}%
\definecolor{mycolor2}{rgb}{0.92900,0.69400,0.12500}
\definecolor{mycolor3}{rgb}{0.92900,0.69400,0.12500}%
\definecolor{mycolor4}{rgb}{0.49400,0.18400,0.55600}%
\definecolor{mycolor4}{rgb}{0.46600,0.67400,0.18800}
\begin{tikzpicture}

\begin{axis}[%
width=0.6\linewidth,
height = 0.4\linewidth,
scale only axis,
bar shift auto,
xmin=0.509090909090909,
xmax=2.48571428571429,
xmax=2.48571428571429,
xtick=data,
xticklabels={0 dB,10 dB},
ymin=0,
ymax=0,
ymax=30,
xmajorgrids,
ymajorgrids,
title style={font=\bfseries},
ylabel={$\text{P}_{\text{TX}}\text{ [dBm]}$},
axis background/.style={fill=white},
legend style={legend cell align=left, align=left, draw=white!15!black, at = {(0.9,0.3)}}
]
\addplot[ybar, bar width=10, draw=mycolor1,fill=mycolor1, area legend] table[row sep=crcr] {%
1 17.7435771443534 \\
2 28.60144809083596\\
};
\addlegendentry{Proposed algorithm}

\addplot[ybar, bar width=10, draw=mycolor2,fill=mycolor2, area legend] table[row sep=crcr] {%
1 17.8569387590180\\
2 28.65343030404450\\
};
\addlegendentry{Algorithm in \cite{wu2019intelligent}}

\end{axis}

\end{tikzpicture}%
    \caption{Comparison between the proposed algorithm and the benchmark in \cite{wu2019intelligent} in terms of required transmit power to achieve a target SINR of $0$ and $10 ~\deci\bel$ with 6 UEs.}
    \label{fig:ZhangComparisonBar}
\end{figure}
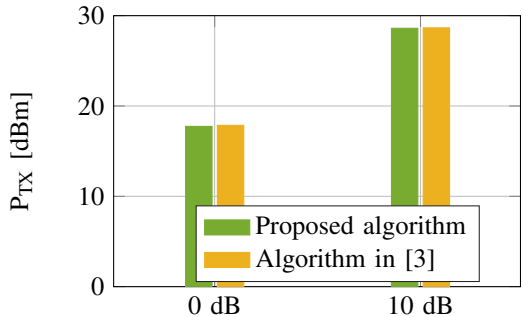
The first comparison is performed w.r.t. \cite{wu2019intelligent} and is reported in Figure \ref{fig:ZhangComparisonBar}. In this case, to handle the high computational complexity of the algorithm in \cite{wu2019intelligent} (see Section \ref{sec:computational_complexity}), a simplified system setup is adopted, assuming that each RIS is equipped with only $40$ scattering elements. 
{As reported in Figure \ref{fig:ZhangComparisonBar}, the two algorithms provide nearly identical results in terms of required transmitted power. However, our approach offers a significant reduction in complexity, as discussed in Section \ref{sec:computational_complexity}.} 

\begin{figure}
    \centering
%
\definecolor{mycolor2}{rgb}{0.00000,0.44700,0.74100}%
\definecolor{mycolor1}{rgb}{0.46600,0.67400,0.18800}%
\definecolor{mycolor2}{rgb}{0.92900,0.69400,0.12500}
\definecolor{mycolor4}{rgb}{0.49400,0.18400,0.55600}%
\definecolor{mycolor3}{rgb}{0.85000,0.32500,0.09800}
\begin{tikzpicture}
\begin{axis}[%
width=0.6\linewidth,
height = 0.3\linewidth,
scale only axis,
xmin=0,
xmax=150,
xlabel style={font=\color{white!15!black}},
ymin=16,
ymax=28,
ylabel style={font=\color{white!15!black}},
ylabel={$\text{P}_{\text{TX}}\text{[dBm]}$},
axis background/.style={fill=white},
title style={font=\bfseries},
xmajorgrids,
ymajorgrids,
legend style={legend cell align=left, align=left, draw=white!15!black}
]
\addplot [color=mycolor3, line width = 2.0 pt]
  table[row sep=crcr]{%
1	27.9984358496266\\
2	25.4658467497103\\
3	23.8102128473181\\
4	22.9951964814729\\
5	22.475721399597\\
6	22.0260142251365\\
7	21.6964150485779\\
8	21.3761407791152\\
9	21.113429141068\\
10	20.8736741640214\\
11	20.6406040071119\\
12	20.4655809925897\\
13	20.2768924357422\\
14	20.1000014760159\\
15	19.9271689572587\\
16	19.7819510986243\\
17	19.6611030670001\\
18	19.512590302226\\
19	19.3778443034676\\
20	19.2843373670809\\
21	19.1486152961543\\
22	19.0255314370083\\
23	18.9727459687838\\
24	18.8742433239113\\
25	18.8010129678044\\
26	18.7214657123135\\
27	18.6433823866816\\
28	18.5738085842111\\
29	18.5006352695428\\
30	18.4341938489791\\
31	18.3661045845821\\
32	18.3107390384373\\
33	18.2510777572468\\
34	18.2070524115445\\
35	18.1553759561044\\
36	18.1137947140371\\
37	18.0735630618152\\
38	18.0399746778399\\
39	17.9940468487175\\
40	17.9612167981408\\
41	17.9177984681104\\
42	17.8864093380334\\
43	17.8467087440249\\
44	17.814696964792\\
45	17.785949036657\\
46	17.7533615937089\\
47	17.7222976653304\\
48	17.6880622924738\\
49	17.6558054124455\\
50	17.6251439336574\\
51	17.5960670492269\\
52	17.5744274974747\\
53	17.5524327366044\\
54	17.5280418511474\\
55	17.5121996231567\\
56	17.5020778585915\\
57	17.4927886314432\\
58	17.4831457033409\\
59	17.4749467173791\\
60	17.4653683450824\\
61	17.4571213501265\\
62	17.4490469793374\\
63	17.4405107344369\\
64	17.431797409664\\
65	17.4246099181859\\
66	17.4169449998148\\
67	17.4098858719955\\
68	17.4009925027483\\
69	17.3944921190289\\
70	17.3879864781626\\
71	17.3807612910209\\
72	17.373760369079\\
73	17.3668238786158\\
74	17.3603966382016\\
75	17.3544925402345\\
76	17.3476886584955\\
77	17.3412153994026\\
78	17.3354028020142\\
79	17.3305936997649\\
80	17.3266243130891\\
81	17.3223147803818\\
82	17.3179742183161\\
83	17.3140562201895\\
84	17.3099310559211\\
85	17.3036655442117\\
86	17.3041348167842\\
87	17.3041348167842\\
88	17.3041348167842\\
89	17.3041348167842\\
90	17.3041348167842\\
91	17.3041348167842\\
92	17.3041348167842\\
93	17.3041348167842\\
94	17.3041348167842\\
95	17.3041348167842\\
96	17.3041348167842\\
97	17.3041348167842\\
98	17.3041348167842\\
99	17.3041348167842\\
100	17.3041348167842\\
101	17.3041348167842\\
102	17.3041348167842\\
103	17.3041348167842\\
104	17.3041348167842\\
105	17.3041348167842\\
106	17.3041348167842\\
107	17.3041348167842\\
108	17.3041348167842\\
109	17.3041348167842\\
110	17.3041348167842\\
111	17.3041348167842\\
112	17.3041348167842\\
113	17.3041348167842\\
114	17.3041348167842\\
115	17.3041348167842\\
116	17.3041348167842\\
117	17.3041348167842\\
118	17.3041348167842\\
119	17.3041348167842\\
120	17.3041348167842\\
121	17.3041348167842\\
122	17.3041348167842\\
123	17.3041348167842\\
124	17.3041348167842\\
125	17.3041348167842\\
126	17.3041348167842\\
127	17.3041348167842\\
128	17.3041348167842\\
129	17.3041348167842\\
130	17.3041348167842\\
131	17.3041348167842\\
132	17.3041348167842\\
133	17.3041348167842\\
134	17.3041348167842\\
135	17.3041348167842\\
136	17.3041348167842\\
137	17.3041348167842\\
138	17.3041348167842\\
139	17.3041348167842\\
140	17.3041348167842\\
141	17.3041348167842\\
142	17.3041348167842\\
143	17.3041348167842\\
144	17.3041348167842\\
145	17.3041348167842\\
146	17.3041348167842\\
147	17.3041348167842\\
148	17.3041348167842\\
149	17.3041348167842\\
150	17.3041348167842\\
};
\addlegendentry{SCA approach \cite{kumar2023novel}, case 1}

\addplot [color=mycolor1, line width = 2.0 pt]
  table[row sep=crcr]{%
1	27.9984356540492\\
2	18.8558395114927\\
3	18.0260914094733\\
4	17.8275836660085\\
5	17.7605523669292\\
6	17.7131882550533\\
7	17.6904320120771\\
8	17.6740235323395\\
9	17.6637345675502\\
10	17.6637345675502\\
11	17.6637345675502\\
12	17.6637345675502\\
13	17.6637345675502\\
14	17.6637345675502\\
15	17.6637345675502\\
16	17.6637345675502\\
17	17.6637345675502\\
18	17.6637345675502\\
19	17.6637345675502\\
20	17.6637345675502\\
21	17.6637345675502\\
22	17.6637345675502\\
23	17.6637345675502\\
24	17.6637345675502\\
25	17.6637345675502\\
26	17.6637345675502\\
27	17.6637345675502\\
28	17.6637345675502\\
29	17.6637345675502\\
30	17.6637345675502\\
31	17.6637345675502\\
32	17.6637345675502\\
33	17.6637345675502\\
34	17.6637345675502\\
35	17.6637345675502\\
36	17.6637345675502\\
37	17.6637345675502\\
38	17.6637345675502\\
39	17.6637345675502\\
40	17.6637345675502\\
41	17.6637345675502\\
42	17.6637345675502\\
43	17.6637345675502\\
44	17.6637345675502\\
45	17.6637345675502\\
46	17.6637345675502\\
47	17.6637345675502\\
48	17.6637345675502\\
49	17.6637345675502\\
50	17.6637345675502\\
51	17.6637345675502\\
52	17.6637345675502\\
53	17.6637345675502\\
54	17.6637345675502\\
55	17.6637345675502\\
56	17.6637345675502\\
57	17.6637345675502\\
58	17.6637345675502\\
59	17.6637345675502\\
60	17.6637345675502\\
61	17.6637345675502\\
62	17.6637345675502\\
63	17.6637345675502\\
64	17.6637345675502\\
65	17.6637345675502\\
66	17.6637345675502\\
67	17.6637345675502\\
68	17.6637345675502\\
69	17.6637345675502\\
70	17.6637345675502\\
71	17.6637345675502\\
72	17.6637345675502\\
73	17.6637345675502\\
74	17.6637345675502\\
75	17.6637345675502\\
76	17.6637345675502\\
77	17.6637345675502\\
78	17.6637345675502\\
79	17.6637345675502\\
80	17.6637345675502\\
81	17.6637345675502\\
82	17.6637345675502\\
83	17.6637345675502\\
84	17.6637345675502\\
85	17.6637345675502\\
86	17.6637345675502\\
87	17.6637345675502\\
88	17.6637345675502\\
89	17.6637345675502\\
90	17.6637345675502\\
91	17.6637345675502\\
92	17.6637345675502\\
93	17.6637345675502\\
94	17.6637345675502\\
95	17.6637345675502\\
96	17.6637345675502\\
97	17.6637345675502\\
98	17.6637345675502\\
99	17.6637345675502\\
100	17.6637345675502\\
101	17.6637345675502\\
102	17.6637345675502\\
103	17.6637345675502\\
104	17.6637345675502\\
105	17.6637345675502\\
106	17.6637345675502\\
107	17.6637345675502\\
108	17.6637345675502\\
109	17.6637345675502\\
110	17.6637345675502\\
111	17.6637345675502\\
112	17.6637345675502\\
113	17.6637345675502\\
114	17.6637345675502\\
115	17.6637345675502\\
116	17.6637345675502\\
117	17.6637345675502\\
118	17.6637345675502\\
119	17.6637345675502\\
120	17.6637345675502\\
121	17.6637345675502\\
122	17.6637345675502\\
123	17.6637345675502\\
124	17.6637345675502\\
125	17.6637345675502\\
126	17.6637345675502\\
127	17.6637345675502\\
128	17.6637345675502\\
129	17.6637345675502\\
130	17.6637345675502\\
131	17.6637345675502\\
132	17.6637345675502\\
133	17.6637345675502\\
134	17.6637345675502\\
135	17.6637345675502\\
136	17.6637345675502\\
137	17.6637345675502\\
138	17.6637345675502\\
139	17.6637345675502\\
140	17.6637345675502\\
141	17.6637345675502\\
142	17.6637345675502\\
143	17.6637345675502\\
144	17.6637345675502\\
145	17.6637345675502\\
146	17.6637345675502\\
147	17.6637345675502\\
148	17.6637345675502\\
149	17.6637345675502\\
150	17.6637345675502\\
151	17.6637345675502\\
};
\addlegendentry{Proposed approach, case 1}

\end{axis}

\end{tikzpicture}%
%
%
\definecolor{mycolor2}{rgb}{0.00000,0.44700,0.74100}%
\definecolor{mycolor1}{rgb}{0.46600,0.67400,0.18800}%
\definecolor{mycolor2}{rgb}{0.92900,0.69400,0.12500}
\definecolor{mycolor4}{rgb}{0.49400,0.18400,0.55600}%
\definecolor{mycolor3}{rgb}{0.85000,0.32500,0.09800}
\begin{tikzpicture}

\begin{axis}[%
width=0.6\linewidth,
height = 0.3\linewidth,
scale only axis,
xmin=0,
xmax=150,
ymin=15,
ymax=40,
ylabel style={font=\color{white!15!black}},
ylabel={$\text{P}_{\text{TX}}\text{[dBm]}$},
axis background/.style={fill=white},
title style={font=\bfseries},
xmajorgrids,
ymajorgrids,
xlabel={Iterations},
legend style={legend cell align=left, align=left, draw=white!15!black}
]
\addplot [color=mycolor4, line width = 2.0 pt]
  table[row sep=crcr]{%
1	39.3257983492934\\
2	33.25977728391\\
3	27.3064263845876\\
4	24.93601537673\\
5	23.68071275076\\
6	22.771507063426\\
7	22.100620912714\\
8	21.5752327519504\\
9	21.1945633109386\\
10	20.8217049531443\\
11	20.4920846689554\\
12	20.2232005274484\\
13	19.9985628988308\\
14	19.7438167925121\\
15	19.5413718978777\\
16	19.3047384669354\\
17	19.1697572802318\\
18	18.98474251939\\
19	18.8326551503231\\
20	18.7293886226341\\
21	18.5522309691133\\
22	18.4096249196975\\
23	18.297739508915\\
24	18.2128392328366\\
25	18.110798210394\\
26	18.0137049904653\\
27	17.8670879321727\\
28	17.866696160585\\
29	17.8177085748613\\
30	17.7693503100912\\
31	17.7256575699939\\
32	17.682709876829\\
33	17.6173449349026\\
34	17.6065676699561\\
35	17.5948164459001\\
36	17.5838051287373\\
37	17.5725836194856\\
38	17.5621459675841\\
39	17.5517660970712\\
40	17.5416081759173\\
41	17.5320517235565\\
42	17.5223479371544\\
43	17.5129003247238\\
44	17.5048336456249\\
45	17.4953759504127\\
46	17.4867805699279\\
47	17.4784874303135\\
48	17.4702276706589\\
49	17.4622272480434\\
50	17.454428662304\\
51	17.4468097242961\\
52	17.4393536315378\\
53	17.4318766034864\\
54	17.425431707022\\
55	17.4181337619659\\
56	17.4113651655222\\
57	17.4046024043817\\
58	17.3980324143612\\
59	17.3916585980209\\
60	17.3852878911481\\
61	17.37935067214\\
62	17.3731530527555\\
63	17.3673535004667\\
64	17.3614982946906\\
65	17.3558033796157\\
66	17.3501737580346\\
67	17.3447074074777\\
68	17.3393823120938\\
69	17.3339732535215\\
70	17.3291720983174\\
71	17.3238404114815\\
72	17.3186697701425\\
73	17.3137093220956\\
74	17.308876040568\\
75	17.3040717504667\\
76	17.2993253764423\\
77	17.2948107897752\\
78	17.2901756493951\\
79	17.2856864786763\\
80	17.2813534878086\\
81	17.2770155127831\\
82	17.2726994701777\\
83	17.2684377452255\\
84	17.2643270658386\\
85	17.2601443439989\\
86	17.256098517359\\
87	17.2520238008354\\
88	17.248357923508\\
89	17.2443509366488\\
90	17.240507940654\\
91	17.2367455912042\\
92	17.233046453073\\
93	17.2293604843603\\
94	17.225726888767\\
95	17.2221896739735\\
96	17.2187152217239\\
97	17.2152010966398\\
98	17.2117731891217\\
99	17.2084673201669\\
100	17.2050404103831\\
101	17.2017175684615\\
102	17.1984585761932\\
103	17.1952587253748\\
104	17.1920629098936\\
105	17.1882744606496\\
106	17.186377671645\\
107	17.1825915750006\\
108	17.1798937252955\\
109	17.1768842038659\\
110	17.1737332437649\\
111	17.1709529149203\\
112	17.1679485177501\\
113	17.1650603116437\\
114	17.1622077008862\\
115	17.1593632659149\\
116	17.1566313708515\\
117	17.1538962957318\\
118	17.1511703025797\\
119	17.1486090879943\\
120	17.1458764049708\\
121	17.1431915965333\\
122	17.1405868867565\\
123	17.137935398282\\
124	17.1355460935226\\
125	17.1330106465067\\
126	17.1309719560643\\
127	17.1280602852185\\
128	17.1255735198728\\
129	17.1230989046408\\
130	17.1206330407226\\
131	17.1184102726888\\
132	17.1161006477149\\
133	17.1139553114\\
134	17.1113549483906\\
135	17.1090381281023\\
136	17.1066771443753\\
137	17.1046060262223\\
138	17.1022947525471\\
139	17.1001794127116\\
140	17.0979669023293\\
141	17.0957714883073\\
142	17.0936199601757\\
143	17.0914923522897\\
144	17.0893553063728\\
145	17.0873419815766\\
146	17.085662737826\\
147	17.083296209798\\
148	17.0811994327026\\
149	17.0791810657352\\
150	17.0771283171464\\
};
\addlegendentry{SCA approach \cite{kumar2023novel}, case 2}

\addplot [color=mycolor2, line width = 2.0 pt]
  table[row sep=crcr]{%
1	39.3257151100176\\
2	20.2642635778249\\
3	17.9940403679733\\
4	16.8339498742193\\
5	16.5403183290271\\
6	16.383145404699\\
7	16.3201598486279\\
8	16.3201598486279\\
9	16.3201598486279\\
10	16.3201598486279\\
11	16.3201598486279\\
12	16.3201598486279\\
13	16.3201598486279\\
14	16.3201598486279\\
15	16.3201598486279\\
16	16.3201598486279\\
17	16.3201598486279\\
18	16.3201598486279\\
19	16.3201598486279\\
20	16.3201598486279\\
21	16.3201598486279\\
22	16.3201598486279\\
23	16.3201598486279\\
24	16.3201598486279\\
25	16.3201598486279\\
26	16.3201598486279\\
27	16.3201598486279\\
28	16.3201598486279\\
29	16.3201598486279\\
30	16.3201598486279\\
31	16.3201598486279\\
32	16.3201598486279\\
33	16.3201598486279\\
34	16.3201598486279\\
35	16.3201598486279\\
36	16.3201598486279\\
37	16.3201598486279\\
38	16.3201598486279\\
39	16.3201598486279\\
40	16.3201598486279\\
41	16.3201598486279\\
42	16.3201598486279\\
43	16.3201598486279\\
44	16.3201598486279\\
45	16.3201598486279\\
46	16.3201598486279\\
47	16.3201598486279\\
48	16.3201598486279\\
49	16.3201598486279\\
50	16.3201598486279\\
51	16.3201598486279\\
52	16.3201598486279\\
53	16.3201598486279\\
54	16.3201598486279\\
55	16.3201598486279\\
56	16.3201598486279\\
57	16.3201598486279\\
58	16.3201598486279\\
59	16.3201598486279\\
60	16.3201598486279\\
61	16.3201598486279\\
62	16.3201598486279\\
63	16.3201598486279\\
64	16.3201598486279\\
65	16.3201598486279\\
66	16.3201598486279\\
67	16.3201598486279\\
68	16.3201598486279\\
69	16.3201598486279\\
70	16.3201598486279\\
71	16.3201598486279\\
72	16.3201598486279\\
73	16.3201598486279\\
74	16.3201598486279\\
75	16.3201598486279\\
76	16.3201598486279\\
77	16.3201598486279\\
78	16.3201598486279\\
79	16.3201598486279\\
80	16.3201598486279\\
81	16.3201598486279\\
82	16.3201598486279\\
83	16.3201598486279\\
84	16.3201598486279\\
85	16.3201598486279\\
86	16.3201598486279\\
87	16.3201598486279\\
88	16.3201598486279\\
89	16.3201598486279\\
90	16.3201598486279\\
91	16.3201598486279\\
92	16.3201598486279\\
93	16.3201598486279\\
94	16.3201598486279\\
95	16.3201598486279\\
96	16.3201598486279\\
97	16.3201598486279\\
98	16.3201598486279\\
99	16.3201598486279\\
100	16.3201598486279\\
101	16.3201598486279\\
102	16.3201598486279\\
103	16.3201598486279\\
104	16.3201598486279\\
105	16.3201598486279\\
106	16.3201598486279\\
107	16.3201598486279\\
108	16.3201598486279\\
109	16.3201598486279\\
110	16.3201598486279\\
111	16.3201598486279\\
112	16.3201598486279\\
113	16.3201598486279\\
114	16.3201598486279\\
115	16.3201598486279\\
116	16.3201598486279\\
117	16.3201598486279\\
118	16.3201598486279\\
119	16.3201598486279\\
120	16.3201598486279\\
121	16.3201598486279\\
122	16.3201598486279\\
123	16.3201598486279\\
124	16.3201598486279\\
125	16.3201598486279\\
126	16.3201598486279\\
127	16.3201598486279\\
128	16.3201598486279\\
129	16.3201598486279\\
130	16.3201598486279\\
131	16.3201598486279\\
132	16.3201598486279\\
133	16.3201598486279\\
134	16.3201598486279\\
135	16.3201598486279\\
136	16.3201598486279\\
137	16.3201598486279\\
138	16.3201598486279\\
139	16.3201598486279\\
140	16.3201598486279\\
141	16.3201598486279\\
142	16.3201598486279\\
143	16.3201598486279\\
144	16.3201598486279\\
145	16.3201598486279\\
146	16.3201598486279\\
147	16.3201598486279\\
148	16.3201598486279\\
149	16.3201598486279\\
150	16.3201598486279\\
151	16.3201598486279\\
};
\addlegendentry{Proposed approach, case 2}

\end{axis}

\end{tikzpicture}%
    \caption{Comparison between the proposed algorithm (for $K = 4$) and the SCA approach in \cite{kumar2023novel} for 2 RISs, 4 UEs, $\kappa_s = \kappa_t = 1$ (case 1) and $\kappa_s = \kappa_t = 10$ (case 2).}
    \label{fig:ConfrontoNAM}
\end{figure}
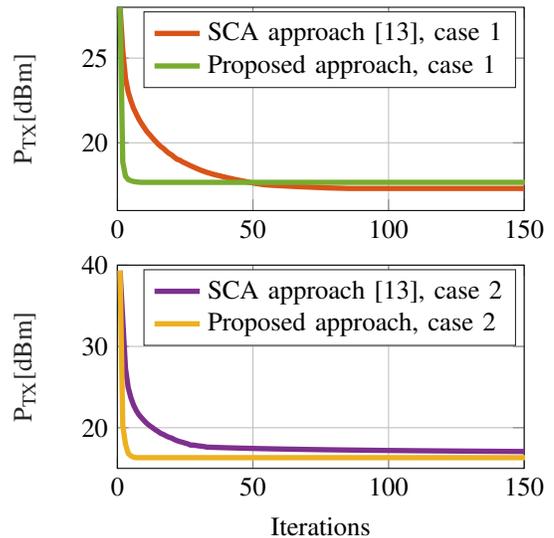

The second comparison is performed w.r.t.\cite{kumar2023novel} which solves the BS precoder and RIS design problem at the same time.
According to the setup considered in \cite{kumar2023novel}, the results are obtained by assuming an operating frequency of $2$ GHz with a bandwidth of 20 MHz. The center of the BS uniform linear array, made of 4 antennas, is located at $[0, 20, 10]$ m. All simulation parameters pertaining to the SCA approach are kept the same as in \cite{kumar2023novel}. We simulated a scenario with $N_u = 4$ users which are assumed to be randomly distributed inside a disk of radius $50$ m, centered at $[100, 100, 2 ]$ m. Similar to \cite{kumar2023novel}, we consider Rician distributed small-scale fading and a distance-dependent path loss for all the wireless links with Rician factors equal to $\kappa_t = \kappa_s = 1$ (case 1) and $10$ (case 2). Two RISs with size $36 \times 16$ are considered, the first located as detailed in \cite{kumar2023novel}, at $[30, 0 , 5 ]$ m, and the second located at $[35, 0 , 5 ]$ m.  
Also, we assumed $K = N_u = 4$. To ensure a fair comparison, the simulations were run multiple times {considering different locations for the users and NLOS components}.
The results, reported in Figure \ref{fig:ConfrontoNAM}, {show that the two algorithms have comparable performance. Notably, the proposed algorithm converges in fewer iterations and has lower computational complexity (see Section \ref{sec:computational_complexity}). The same convergence trend is shown in Figures \ref{fig:changingK}, \ref{fig:changingKsingleTile} \ref{fig:ConSenzaProj} and \ref{fig:IOSMInvertiti}, demonstrating that, in all the considered setups, the proposed AO scheme requires few iterations to reach convergence.}

\section{Conclusion and Future Research}
\label{sec:future}
To tackle the problem of power minimization and system fairness in RIS-aided scenarios, the study presented efficient optimization algorithms to minimize the transmit power subjected to \ac{SINR} constraints in a single-cell multi-user system. The proposed approach optimizes the BS precoder and RIS configuration by using an iterative AO algorithm. The BS precoder is tackled through the \ac{SOCP} method, and the RISs are optimized solving an \ac{MSE} minimization problem. The obtained simulation results have demonstrated the effectiveness of the proposed method, showcasing good performance alongside a notably reduction in computation complexity when compared to recent benchmark schemes available in the literature.

\section*{Appendix A}
\hypertarget{A}{}
{
\begin{proof} Suppose, by way of contradiction, that $\mathbf{V}$ constitutes a solution to \hyperlink{PO}{PO} and that $\text{SINR}_j\left(\boldsymbol{\alpha}^{(t)},\mathbf{V}\right) >\Sigma_j$. Under such circumstances, there always exists a scaled vector $\tilde{\mbf{v}}_{j}$
 \begin{equation}
 \tilde{\mbf{v}}_{j}=\mbf{v}_{j}\frac{\Sigma_j}{\text{SINR}_j\left(\boldsymbol{\alpha}^{(t)},\mathbf{V}\right)}
 \end{equation}
 such that the constraint is met with equality. Since $\frac{\Sigma_j}{\text{SINR}_j\left(\boldsymbol{\alpha}^{(t)},\mathbf{V}\right)}<1$, we have
 \begin{equation}
 \sum\limits_{i = 1,i\ne j}^{N_{u}} \lVert \mbf{v}_{i} \rVert ^2 +\lVert \tilde{\mbf{v}}_{j} \rVert ^2 \le \sum\limits_{i = 1}^{N_{u}} \lVert \mbf{v}_{i} \rVert ^2.
 \end{equation}
 Accordingly, the matrix $\mbf{\tilde{V}}$, obtained from $\mbf{V}$ by replacing the column $j$ with $\mbf{\tilde{v}}_{j}$, is a feasible vector for \hyperlink{PO}{PO} that would consume less power but this actually violates the hypothesis that $\mathbf{V}$ is a solution of \hyperlink{PO}{PO}. \end{proof}}

\section*{Appendix B}
\hypertarget{AppendixB}{}
\begin{algorithm}
\small
    \SetAlgoLined
    \textbf{Initialization for $n=0$;}\\
    -Set initial values of $\boldsymbol{\lambda}^{(0)}$ and $\boldsymbol{\mu}^{(0)}$ that fulfill constraints of \eqref{eq:LagrangianDualProblem};\\
    -Set a suitable step size $\delta$;\\
    -Set an arbitrarily small value $\varepsilon$, $\Delta=1$;\\ 
    \While{$\Delta > \varepsilon$}{
        - \textbf{Apply \ac{GAM}}\\
        - Compute dual function gradients $\nabla_{\boldsymbol{\lambda}}d(\boldsymbol{\lambda}^{(n)},\boldsymbol{\mu}^{(n)})$ and $\nabla_{\boldsymbol{\mu}}d(\boldsymbol{\lambda}^{(n)},\boldsymbol{\mu}^{(n)})$(\eqref{eq:partDerg2}, \eqref{eq:partDerg5}); \\
        - Compute $\boldsymbol{\lambda}^{(n + 1)}$ and $\boldsymbol{\mu}^{(n + 1)}$ (\eqref{eq:LangrangianMultipliersUpdate});\\
        - $\Delta_{\boldsymbol{\lambda}}=    \frac{\lVert\boldsymbol{\lambda}^{(n + 1)} - \boldsymbol{\lambda}^{(n)}\rVert}{\lVert \boldsymbol{\lambda}^{(n)}\rVert};$\\
        - $\Delta_{\boldsymbol{\mu}}=    \frac{\lVert\boldsymbol{\mu}^{(n + 1)} - \boldsymbol{\mu}^{(n)}\rVert}{\lVert \boldsymbol{\mu}^{(n)}\rVert};$\\
        - $\Delta = \max(\Delta_{\boldsymbol{\lambda}}, \Delta_{\boldsymbol{\lambda}})$;\\
        - n = n + 1;\\
        }
Output: $\boldsymbol{\alpha}^\star$\\
\caption{Tile Configuration Optimization}
\label{alg:TilesConfiguration}
\end{algorithm}
{
\subsection{Deriving the Lagrange Dual Function}
The optimization problem \hyperlink{TO2}{TO2} is convex and differentiable and can be solved in the dual domain. 
The Lagrangian function $\mathcal{L}\left(\boldsymbol{\alpha},\boldsymbol{\lambda},\boldsymbol{\mu}\right)$ for the constrained problem \hyperlink{TO2}{TO2} is
\begin{equation}
\label{eq:Lagrangian}
\begin{aligned}  \mathcal{L}\left(\boldsymbol{\alpha},\boldsymbol{\lambda},\boldsymbol{\mu}\right)&=\sum\limits_{i = 1}^{N_{u}}  {E}_{i}\left(g_i;\boldsymbol{\alpha},\mathbf{V}\right)+\sum\limits_{i = 1}^{N_{u}} \lambda_i \left({E}_{i}\left(g_i;\boldsymbol{\alpha},\mathbf{V}\right) - \mathcal{E}_i\right) \\
    &+ \sum\limits_{k = 1}^{K}\mu_{k}\left(\boldsymbol{\alpha}\mathbf{\tilde{Q}}_{k}\boldsymbol{\alpha}^H-P\right)\\
&=\sum\limits_{i = 1}^{N_{u}}(1+\lambda_i)  {E}_{i}\left(g_i;\boldsymbol{\alpha},\mathbf{V}\right) + \sum\limits_{k = 1}^{K}\mu_{k}\boldsymbol{\alpha} \mathbf{\tilde{Q}}_{k}\boldsymbol{\alpha}^H \\
& - \sum\limits_{i = 1}^{N_{u}}\lambda_i\mathcal{E}_i - P\sum\limits_{k = 1}^{K}\mu_{k},\
\end{aligned}
\end{equation}
with $\boldsymbol{\lambda} = \{\lambda_i\}$ and $\boldsymbol{\mu} = \{\mu_{k}\}$ being the vectors of Lagrangian multipliers associated to the minimum \ac{SINR} and the \ac{GC} constraints, respectively.
The Lagrangian dual function is consequently defined as 
\begin{equation}
  \label{eq:lagrangianDualFunct}  d(\boldsymbol{\lambda},\boldsymbol{\mu})=\min\limits_{\boldsymbol{\alpha}} \mathcal{L}\left(\boldsymbol{\alpha},\boldsymbol{\lambda},\boldsymbol{\mu}\right).
\end{equation}
Since both the objective function and the constraints are convex in $\boldsymbol{\alpha}$, $\mathcal{L}\left(\boldsymbol{\alpha},\boldsymbol{\lambda},\boldsymbol{\mu}\right)$ is convex {in} $\boldsymbol{\alpha}$ and, for fixed values of $\boldsymbol{\lambda}$ and $\boldsymbol{\mu}$, the optimal value of $\boldsymbol{\alpha}^\star(\boldsymbol{\lambda},\boldsymbol{\mu})$ that minimizes \eqref{eq:Lagrangian} can be found by setting to zero the gradient of $\mathcal{L}\left(\boldsymbol{\alpha},\boldsymbol{\lambda},\boldsymbol{\mu}\right)$ computed w.r.t. $\boldsymbol{\alpha}$. Exploiting the knowledge of the gradients in \eqref{eq:GradientGC}
and \eqref{eq:GradientMSE} yields 
\begin{equation}
\label{eq:gradLagrangian}
\begin{aligned}   \nabla_{\boldsymbol{\alpha}}\mathcal{L}\left(\boldsymbol{\alpha},\boldsymbol{\lambda},\boldsymbol{\mu}\right)&=\sum\limits_{i = 1}^{N_{u}}  (1+\lambda_i)
    (|g_i|^2\mathbf{\bar{h}}_i\mathbf{T}\mathbf{\tilde{H}}_i^H-g_i\mathbf{v}_i^H\mathbf{\tilde{H}}_i^H)\\ &+\boldsymbol{\alpha}\left(\sum\limits_{i = 1}^{N_{u}}  (1+\lambda_i) |g_i|^2\mathbf{\tilde{H}}_i\mathbf{T}\mathbf{\tilde{H}}_i^H  + \sum\limits_{k = 1}^{K} \mu_{k}\mathbf{\tilde{Q}}_k\right)
\end{aligned}
\end{equation}
The stationary point $\boldsymbol{\alpha}^\star(\boldsymbol{\lambda},\boldsymbol{\mu})$ that sets the gradient to zero is
\begin{equation}
\label{eq:gammaOpt}
\begin{aligned}  \boldsymbol{\alpha}^\star(\boldsymbol{\lambda},\boldsymbol{\mu})&=\sum\nolimits_{i = 1}^{N_{u}}  (1+\lambda_i)
    (g_i\mathbf{v}_i^H\mathbf{\tilde{H}}_i^H-|g_i|^2\mathbf{\bar{h}}_i\mathbf{T}\mathbf{\tilde{H}}_i^H)\\
    &\times\left(\sum\nolimits_{i = 1}^{N_{u}}  (1+\lambda_i)    |g_i|^2\mathbf{\tilde{H}}_i\mathbf{T}\mathbf{\tilde{H}}_i^H + \sum\nolimits_{k = 1}^{K} \mu_{k}\mathbf{\tilde{Q}}_k\right)^{-1}.
    \end{aligned}
\end{equation}
The Lagrangian dual function is computed by replacing $\boldsymbol{\alpha}^\star(\boldsymbol{\lambda},\boldsymbol{\mu})$ in \eqref{eq:Lagrangian}, i.e., 
\begin{equation}
 d(\boldsymbol{\lambda},\boldsymbol{\mu})=  \mathcal{L}\left(\boldsymbol{\alpha}^\star(\boldsymbol{\lambda},\boldsymbol{\mu}),\boldsymbol{\lambda},\boldsymbol{\mu}\right).
\end{equation}
\subsection{Solving the Lagrangian Dual Problem}
Since the primal problem \hyperlink{TO2}{TO2} is strongly dual (please refer to Appendix \hyperlink{AppendixC}{C} for further information and assumptions), its solution can be found by solving the \ac{LDP}
\begin{equation}
\label{eq:LagrangianDualProblem}    \max\limits_{\boldsymbol{\lambda}\geq \mathbf{0},\boldsymbol{\mu}\geq \mathbf{0}} d(\boldsymbol{\lambda},\boldsymbol{\mu}).
\end{equation}
The Lagrangian dual function is differentiable and its gradients $\nabla_{\boldsymbol{\lambda}} d(\boldsymbol{\lambda},\boldsymbol{\mu})$ and $\nabla_{\boldsymbol{\mu}} d(\boldsymbol{\lambda},\boldsymbol{\mu})$ are computed in the following.
Denoting for ease of notation ${E}_{l}(\boldsymbol{\lambda},\boldsymbol{\mu}) = {E}_{l}\left(g_l;\boldsymbol{\alpha}^\star(\boldsymbol{\lambda},\boldsymbol{\mu}),\mathbf{V}\right)$ and $\boldsymbol{\alpha} = \boldsymbol{\alpha}^\star(\boldsymbol{\lambda},\boldsymbol{\mu})$, for a generic $ l = 1, \ldots, N_u$, from \eqref{eq:Lagrangian} we get
\begin{equation}
\begin{aligned}
\label{eq:partDerg2}
    \frac{\partial d(\boldsymbol{\lambda},\boldsymbol{\mu})}{\partial \lambda_l} & = {E}_{l}(\boldsymbol{\lambda},\boldsymbol{\mu})-\mathcal{E}_l+\sum\nolimits_{i = 1}^{N_{u}}(1+\lambda_i)\frac{\partial {E}_{i}(\boldsymbol{\lambda},\boldsymbol{\mu})}{\partial \lambda_l}\\
    & +  2\Re\left[\frac{\partial \boldsymbol{\alpha}}{\partial \lambda_l}\sum\nolimits_{k = 1}^{K}\mu_{k} \mathbf{\tilde{Q}}_{k}
    \boldsymbol{\alpha}^H\right]
\end{aligned}
\end{equation}
where, by replacing the value in \eqref{eq:MSEComp} for the MSE $E_i$, we obtain
\begin{equation}
\label{eq:partDer}
\begin{aligned}
    \frac{\partial {E}_{i}(\boldsymbol{\lambda},\boldsymbol{\mu})}{\partial \lambda_l}=&-\frac{\partial \boldsymbol{\alpha}}{\partial \lambda_l}\left(g_i^*\mathbf{\tilde{H}}_i\mathbf{v}_i-|g_i|^2\mathbf{\tilde{H}}_i\mathbf{T}\mathbf{\bar{h}}_i^H\right)\\
    &-\left(g_i\mathbf{v}_i^H\mathbf{\tilde{H}}_i^H-|g_i|^2\mathbf{\bar{h}}_i\mathbf{T}\mathbf{\tilde{H}}_i^H\right)\frac{\partial \boldsymbol{\alpha}^H}{\partial \lambda_l}\\
    &+|g_i|^2\left(\frac{\partial \boldsymbol{\alpha}}{\partial \lambda_l}\mathbf{\tilde{H}}_i\mathbf{T}\mathbf{\tilde{H}}_i^H\boldsymbol{\alpha}^H+\boldsymbol{\alpha}\mathbf{\tilde{H}}_i\mathbf{T}\mathbf{\tilde{H}}_i^H\frac{\partial \boldsymbol{\alpha}^H}{\partial \lambda_l}\right).
\end{aligned}
\end{equation}
To compute the partial derivative in \eqref{eq:partDer} we rewrite $\boldsymbol{\alpha}$ as
\begin{equation}
 \boldsymbol{\alpha}=\sum\nolimits_{i = 1}^{N_{u}}(1+\lambda_i)\mathbf{f}_i\left[\sum\nolimits_{i = 1}^{N_{u}}(1+\lambda_i)\mathbf{A}_i+ \Pi\right]^{-1},  
\end{equation}
with 
\begin{equation}
\begin{aligned}
    &\mathbf{f}_i = g_i\mathbf{v}_i^H\mathbf{\tilde{H}}_i^H-|g_i|^2\mathbf{\bar{h}}_i\mathbf{T}\mathbf{\tilde{H}}_i^H\\
   &\mathbf{A}_i = |g_i|^2\mathbf{\tilde{H}}_i\mathbf{T}\mathbf{\tilde{H}}_i^H \\
   &\Pi = \sum\nolimits_{k = 1}^K \mu_{k} \mathbf{\tilde{Q}}_{k}.\\
\end{aligned}
\end{equation}
Accordingly, we can write
\begin{equation}
\label{eq:partDerGam}
\begin{aligned}
    \frac{\partial \boldsymbol{\alpha}}{\partial \lambda_l}= & ~\mathbf{f}_l\left[\sum\nolimits_{i = 1}^{N_{u}}(1+\lambda_i)\mathbf{A}_i\right]^{-1}+ \\ & \sum\nolimits_{i = 1}^{N_{u}}{(1+\lambda_i)}\mathbf{f}_i
   \frac{\partial}{\partial \lambda_l} \left[\sum\nolimits_{i = 1}^{N_{u}}(1+\lambda_i)\mathbf{A}_i+\Pi\right]^{-1}.
\end{aligned}
\end{equation}
By exploiting the identity
\begin{equation}
\label{eq:partDerFin}
\begin{aligned}
     \frac{\partial}{\partial \lambda_l}& \left\{\left[\sum\nolimits_{i = 1}^{N_{u}}(1+\lambda_i)\mathbf{A}_i+\Pi\right]\left[\sum\nolimits_{i = 1}^{N_{u}}(1+\lambda_i)\mathbf{A}_i+\Pi\right]^{-1}\right\}\\
     &=\mathbf{A}_l\left[\sum\nolimits_{i = 1}^{N_{u}}(1+\lambda_i)\mathbf{A}_i+\Pi\right]^{-1}+\left[\sum\nolimits_{i = 1}^{N_{u}}(1+\lambda_i)\mathbf{A}_i+\Pi\right]\\
     &\times \frac{\partial}{\partial \lambda_l} \left[\sum\nolimits_{i = 1}^{N_{u}}(1+\lambda_i)\mathbf{A}_i+\Pi\right]^{-1}=0,
\end{aligned}
\end{equation}
the second term of the right-hand side of \eqref{eq:partDer} can be computed as 
\begin{equation}
\label{eq:partDer4}
\begin{aligned}
 \frac{\partial}{\partial \lambda_l}& \left[\sum\nolimits_{i = 1}^{N_{u}}(1+\lambda_i)\mathbf{A}_i+\Pi\right]^{-1}\\&=-\left[\sum\nolimits_{i = 1}^{N_{u}}(1+\lambda_i)\mathbf{A}_i+\Pi\right]^{-1}\mathbf{A}_l\left[\sum\nolimits_{i = 1}^{N_{u}}(1+\lambda_i)\mathbf{A}_i+\Pi\right]^{-1}.  
\end{aligned}
\end{equation}
By replacing \eqref{eq:partDer4} and \eqref{eq:partDerGam} in \eqref{eq:partDer} and \eqref{eq:partDerg2} we can compute the partial derivative of $d(\boldsymbol{\lambda},\boldsymbol{\mu})$ w.r.t. $\boldsymbol{\lambda}$ and get $\nabla_{\boldsymbol{\lambda}}d(\boldsymbol{\lambda},\boldsymbol{\mu})$. 
Similarly, we can compute the partial derivative of $d(\boldsymbol{\lambda}, \boldsymbol{\mu})$ w.r.t. $\boldsymbol{\mu}$ to get $\nabla_{\boldsymbol{\mu}}d(\boldsymbol{\lambda},\boldsymbol{\mu})$. To elaborate, for generic $r = 1, \ldots, K$ from \eqref{eq:Lagrangian} we get
\begin{equation}
\begin{aligned}
\label{eq:partDerg5}
    \frac{\partial d(\boldsymbol{\lambda},\boldsymbol{\mu})}{\partial \mu_{r}} & = \sum\nolimits_{i = 1}^{N_{u}}(1+\lambda_i)\frac{\partial {E}_{i}(\boldsymbol{\lambda},\boldsymbol{\mu})}{\partial \mu_{r}} + \boldsymbol{\alpha} \mathbf{\tilde{Q}}_{r} \boldsymbol{\alpha}^H - 1 \\
    & + 2\Re\left[\frac{\partial \boldsymbol{\alpha}}{\partial \mu_{r}}\sum\nolimits_{k = 1}^{K}\mu_{k} \mathbf{\tilde{Q}}_{k}\boldsymbol{\alpha}^H\right],
  \end{aligned}
\end{equation}
where, by replacing the value in \eqref{eq:MSEComp} for the MSE $E_i$, we obtain
\begin{equation}
\label{eq:partDer6}
\begin{aligned}
    \frac{\partial {E}_{i}(\boldsymbol{\lambda},\boldsymbol{\mu})}{\partial \mu_{r}}=&-\frac{\partial \boldsymbol{\alpha}}{\partial \mu_{r}}\left(g_i^*\mathbf{\tilde{H}}_i\mathbf{v}_i-|g_i|^2\mathbf{\tilde{H}}_i\mathbf{T}\mathbf{\bar{h}}_i^H\right)\\
    &-\left(g_i\mathbf{v}_i^H\mathbf{\tilde{H}}_i^H-|g_i|^2\mathbf{\bar{h}}_i\mathbf{T}\mathbf{\tilde{H}}_i^H\right)\frac{\partial \boldsymbol{\alpha}^H}{\partial \mu_{r}}\\
    &+|g_i|^2\left(\frac{\partial \boldsymbol{\alpha}}{\partial \mu_{r}}\mathbf{\tilde{H}}_i\mathbf{T}\mathbf{\tilde{H}}_i^H\boldsymbol{\alpha}^H+\boldsymbol{\alpha}\mathbf{\tilde{H}}_i\mathbf{T}\mathbf{\tilde{H}}_i^H\frac{\partial \boldsymbol{\alpha}^H}{\partial \mu_{r}}\right).
\end{aligned}
\end{equation}
To compute the partial derivative in \eqref{eq:partDer6}, we rewrite $\boldsymbol{\alpha}$ from \eqref{eq:gammaOpt} as
\begin{equation}
\boldsymbol{\alpha}=\mathbf{f}\left[\mathbf{A}+ \sum\nolimits_{k = 1}^{K}\mu_{k}\mathbf{\tilde{Q}}_{k}\right]^{-1},  
\end{equation}
with 
\begin{equation}
    \begin{aligned}
         &\mathbf{f} = \sum\nolimits_{i = 1}^{N_{u}}  (1+\lambda_i)
    (g_i\mathbf{v}_i^H\mathbf{\tilde{H}}_i^H-|g_i|^2\mathbf{\bar{h}}_i\mathbf{T}\mathbf{\tilde{H}}_i^H)\\
   &\mathbf{A} = \sum\nolimits_{i = 1}^{N_{u}}  (1+\lambda_i)|g_i|^2\mathbf{\tilde{H}}_i\mathbf{T}\mathbf{\tilde{H}}_i^H. \\   
    \end{aligned}
\end{equation}
Accordingly,
we can write
\begin{equation}
\label{eq:partDerGam2}
\begin{aligned}
    \frac{\partial \boldsymbol{\alpha}}{\partial \mu_{r}}= \mathbf{f} 
   \frac{\partial}{\partial \mu_{r}} \left[\mathbf{A}+ \sum\nolimits_{k = 1}^{K}\mu_{k} \mathbf{\tilde{Q}}_{k}\right]^{-1}.
\end{aligned}
\end{equation}
Similarly, it is easy to derive
\begin{equation}
\label{eq:partDerGam3}
\begin{aligned}
    & \frac{\partial}{\partial \mu_{r}} \left[\mathbf{A}+ \sum\limits_{k = 1}^{K}\mu_{k}\mathbf{\tilde{Q}}_{k}\right]^{-1}  =  -\left[\mathbf{A}+ \sum\limits_{k = 1}^{K}\mu_{k}\mathbf{\tilde{Q}}_{k}\right]^{-1} \mathbf{\tilde{Q}}_{r} \left[\sum\limits_{k = 1}^{K}\mu_{k}\mathbf{\tilde{Q}}_{k}\right]^{-1}
\end{aligned}
\end{equation}
Since it is not possible to find the zeros of the gradients in closed form, we employ the iterative {\ac{GAM}}, to solve \eqref{eq:LagrangianDualProblem}. At step $n+1$, the iterative update of the Langrangian variables is
\begin{equation}
\begin{aligned}
  \label{eq:LangrangianMultipliersUpdate}  \boldsymbol{\lambda}^{(n+1)}=\left(\boldsymbol{\lambda}^{(n)}+\delta \nabla_{\boldsymbol{\lambda}}d\left(\boldsymbol{\lambda}^{(n)},\boldsymbol{\mu}^{(n)}\right)\right)^+\\
     \boldsymbol{\mu}^{(n+1)}=\left(\boldsymbol{\mu}^{(n)}+\delta \nabla_{\boldsymbol{\mu}}d\left(\boldsymbol{\lambda}^{(n)},\boldsymbol{\mu}^{(n)}\right)\right)^+\\
    \end{aligned}
\end{equation}
where $\delta$ is a suitable {step size}.
Once the optimal $\boldsymbol{\lambda}^\star$ and $\boldsymbol{\mu}^\star$ are found, we exploit strong duality (please refer to Appendix \hyperlink{AppendixC}{C} for further information and assumptions) to find the solution of the primal problem by replacing $\boldsymbol{\lambda}^\star$ and $\boldsymbol{\mu}^\star$ in \eqref{eq:gammaOpt}. The complete algorithm is outlined in Algorithm \ref{alg:TilesConfiguration}.}

\section*{Appendix C}
\hypertarget{AppendixC}{}
\label{app:optSolution}   
{
Consider the general optimization problem 
\begin{equation}\label{eq:propProb}
\begin{aligned}
& \min \limits_{x} f(x) & \quad\\
\text{s. t.:} \\
& l_i(x)\le0, \quad  i=1,\dots,r.
\end{aligned}    
\end{equation}
For any feasible point $x$, we denote the set of active inequality constrains as 
\begin{equation}
    A(x)=\{j|l_j(x)=0\}.
\end{equation}
\textbf{Definition 1}. \emph{Let $x^\star$ be a solution of \eqref{eq:propProb}, $x^\star$ is a regular point if the gradients $\nabla l_i(x^\star)$ of the active constraints, i.e., $i\in A(x)$, are linearly independent.}
\textbf{Proposition 1}\cite{bertsekas2016nonlinear}. \emph{If \eqref{eq:propProb} is a convex problem and $x^\star$ is a regular point, then strong duality holds.} 
\begin{proof}
    See \cite{bertsekas2016nonlinear}.
\end{proof}
According to Proposition 1, to prove the strong duality of \hyperlink{TO2}{TO2} we need to show that any solution is a \emph{regular} point, i.e., the gradients of the inequality constraints are linearly independent.  
The gradient of the $K$ \ac{GC} constraints \eqref{eq:conv_RIS} is
\begin{equation}
\label{eq:GradientGC}
    \nabla_{\boldsymbol{\alpha}}\boldsymbol{\alpha} \mathbf{\tilde{Q}}_{k}\boldsymbol{\alpha}^H =\boldsymbol{\alpha} \mathbf{\tilde{Q}}_{k}.
\end{equation}
Computing the $N_u$ gradients of the \ac{MSE} constraints \eqref{eq:convMinMSE_SINR} reduces to calculate the gradients of the MSE ${E}_{i}\left(g_i;\boldsymbol{\alpha},\mathbf{V}\right)$.
Replacing $\mathbf{h}_{i}\left(\boldsymbol{\alpha}\right)$ in  \eqref{eq:MSEmat1} with the expression in \eqref{eq:totalChannel} yields 
\begin{equation}
\label{eq:MSEComp}
\begin{aligned}   {E}_{i}\left(g_i;\boldsymbol{\alpha},\mathbf{V}\right)=&1-g_i^H(\mathbf{\bar{h}}_i+\boldsymbol{\alpha}\mathbf{\tilde{H}}_i)\mathbf{v}_i-g_i\mathbf{v}_i^H(\mathbf{\bar{h}}_i^H+\mathbf{\tilde{H}}_i^H\boldsymbol{\alpha}^H)\\   &+|g_i|^2(\mathbf{\bar{h}}_i\mathbf{T}\mathbf{\bar{h}}_i^H+\mathbf{\bar{h}}_i\mathbf{T}\mathbf{\tilde{H}}_i^H\boldsymbol{\alpha}^H+\boldsymbol{\alpha}\mathbf{\tilde{H}}_i\mathbf{T}\mathbf{\bar{h}}_i^H\\ &+\boldsymbol{\alpha}\mathbf{\tilde{H}}_i\mathbf{T}\mathbf{\tilde{H}}_i^H\boldsymbol{\alpha}^H+\sigma^2_i),
\end{aligned}
\end{equation}
where $\mathbf{T}=\sum\nolimits_{j = 1}^{N_{u}}\mathbf{v}_j\mathbf{v}_j^H$. Computing the $N_u$ gradients yields the $KN_u$-dimensional row vectors
\begin{equation}
\label{eq:GradientMSE}
\nabla_{\boldsymbol{\alpha}}{E}_{i}\left(g_i;\boldsymbol{\alpha},\mathbf{V}\right)=|g_i|^2\mathbf{\bar{h}}_i\mathbf{T}\mathbf{\tilde{H}}_i^H+\boldsymbol{\alpha}|g_i|^2\mathbf{\tilde{H}}_i\mathbf{T}\mathbf{\tilde{H}}_i^H-g_i\mathbf{v}_i^H\mathbf{\tilde{H}}_i^H.
\end{equation}
While a rigorous mathematical justification of the linear independence of the constraint gradients may not be feasible, we provide an empirical sketch of the proof, which effectively conveys the intuition behind the linear independence of these gradients. Such intuition is corroborated by the simulations presented as an example in Figure \ref{fig:ConvergenceLagrangianDual}. 
By design, each matrix $\mathbf{\tilde{Q}}_{k}$ is structured as a block matrix, predominantly comprising zeros except for a square block of size $N_u\times N_u$ that is positioned along the main diagonal. This block corresponds to entries associated with the $k$-th tile. Similarly, the gradient $\boldsymbol{\alpha} \mathbf{\tilde{Q}}_{k}$ manifests as a vector of zeros, except for the elements aligned with the $k$-th tile. This unique configuration ensures that the gradients \eqref{eq:GradientGC} across all the $K$ tiles are mutually linearly independent.
Moreover, given that the gradients \eqref{eq:GradientMSE} predominantly hinge on the propagation channels of the users, and these channels are mutually independent, it is reasonable to infer that the $K+N_u$ gradients, each comprising $KN_u$ elements, remain linearly independent. Consequently, also considering that the linear independence is required only for the subset of the active constraints, any solution of \eqref{eq:MSEComp} is a regular point and strong duality holds for \hyperlink{TO2}{TO2}.}

\bibliographystyle{IEEEtran}
\bibliography{references_short}

\begin{thebibliography}{10}
\providecommand{\url}[1]{#1}
\csname url@samestyle\endcsname
\providecommand{\newblock}{\relax}
\providecommand{\bibinfo}[2]{#2}
\providecommand{\BIBentrySTDinterwordspacing}{\spaceskip=0pt\relax}
\providecommand{\BIBentryALTinterwordstretchfactor}{4}
\providecommand{\BIBentryALTinterwordspacing}{\spaceskip=\fontdimen2\font plus
\BIBentryALTinterwordstretchfactor\fontdimen3\font minus \fontdimen4\font\relax}
\providecommand{\BIBforeignlanguage}[2]{{%
\expandafter\ifx\csname l@#1\endcsname\relax
\typeout{** WARNING: IEEEtran.bst: No hyphenation pattern has been}%
\typeout{** loaded for the language `#1'. Using the pattern for}%
\typeout{** the default language instead.}%
\else
\language=\csname l@#1\endcsname
\fi
#2}}
\providecommand{\BIBdecl}{\relax}
\BIBdecl

\bibitem{direnzo2020smart}
M.~{Di Renzo}, A.~Zappone, M.~Debbah, M.-S. Alouini, C.~Yuen, J.~de~Rosny, and S.~Tretyakov, ``Smart radio environments empowered by reconfigurable intelligent surfaces: How it works, state of research, and the road ahead,'' \emph{IEEE J. Sel. Areas Commun.}, vol.~38, no.~11, pp. 2450--2525, 2020.

\bibitem{fotock2024energy}
R.~K. Fotock, A.~Zappone, and M.~{Di Renzo}, ``Energy efficiency optimization in {RIS}-aided wireless networks: Active versus nearly-passive {RIS} with global reflection constraints,'' \emph{IEEE Trans. Commun.}, vol.~72, no.~1, pp. 257--272, 2024.

\bibitem{wu2019intelligent}
Q.~Wu and R.~Zhang, ``Intelligent reflecting surface enhanced wireless network via joint active and passive beamforming,'' \emph{IEEE Trans. Wireless Commun.}, vol.~18, no.~11, pp. 5394--5409, 2019.

\bibitem{wu2020transmit}
J.~Wu and B.~Shim, ``Transmit power minimization in intelligent reflecting surfaces-aided uplink communications,'' in \emph{2020 IEEE REGION 10 CONFERENCE (TENCON)}, 2020, pp. 108--112.

\bibitem{wu2021power}
------, ``Power minimization of intelligent reflecting surface-aided uplink iot networks,'' in \emph{2021 IEEE Wireless Communications and Networking Conference (WCNC)}, 2021, pp. 1--6.

\bibitem{wang2022transmit}
Y.~Wang, P.~Guan, H.~Yu, and Y.~Zhao, ``Transmit power optimization of simultaneous transmission and reflection {RIS} assisted full-duplex communications,'' \emph{IEEE Access}, vol.~10, pp. 61\,192--61\,200, 2022.

\bibitem{ma2022transmit}
H.~Ma, H.~Wang, H.~Li, and Y.~Feng, ``Transmit power minimization for {STAR}-{RIS}-empowered uplink {NOMA} system,'' \emph{IEEE Wireless Commun. Lett.}, vol.~11, no.~11, pp. 2430--2434, 2022.

\bibitem{zhoy2023transmit}
C.~Zhou, B.~Lyu, Y.~Feng, and D.~T. Hoang, ``Transmit power minimization for {STAR}-{RIS} empowered symbiotic radio communications,'' \emph{IEEE Trans. on Cognitive Commun. and Networking}, vol.~9, no.~6, pp. 1641--1656, 2023.

\bibitem{feng2023resource}
W.~Feng, J.~Tang, Q.~Wu, Y.~Fu, X.~Zhang, D.~K.~C. So, and K.-K. Wong, ``Resource allocation for power minimization in {RIS}-assisted multi-{UAV} networks with {NOMA},'' \emph{IEEE Trans. Commun.}, vol.~71, no.~11, pp. 6662--6676, 2023.

\bibitem{ren2023energy}
H.~Ren, Z.~Zhang, Z.~Peng, L.~Li, and C.~Pan, ``Energy minimization in {RIS}-assisted {UAV}-enabled wireless power transfer systems,'' \emph{IEEE Internet of Things Journal}, vol.~10, no.~7, pp. 5794--5809, 2023.

\bibitem{xie2021joint}
X.~Xie, F.~Fang, and Z.~Ding, ``Joint optimization of beamforming, phase-shifting and power allocation in a multi-cluster {IRS}-{NOMA} network,'' \emph{IEEE Trans. Veh. Technol.}, vol.~70, no.~8, pp. 7705--7717, 2021.

\bibitem{sun2024new}
W.~Sun, S.~Sun, T.~Shi, X.~Su, and R.~Liu, ``A new model of beyond diagonal reconfigurable intelligent surfaces ({BD}-{RIS}) for the corresponding quantization and optimization,'' \emph{IEEE Trans. Wireless Commun.}, 2024.

\bibitem{kumar2023novel}
V.~Kumar, R.~Zhang, M.~{Di Renzo}, and L.-N. Tran, ``A novel {SCA}-based method for beamforming optimization in {IRS}/{RIS}-assisted {MU}-{MISO} downlink,'' \emph{IEEE Wireless Commun. Lett.}, vol.~12, no.~2, pp. 297--301, 2023.

\bibitem{kai2024delay}
C.~Kai, L.~Ma, S.~He, J.~Zhu, and W.~Huang, ``Delay pre-compensation scheme for single carrier {IRS}-aided wideband multi-user {MISO} systems,'' \emph{IEEE Commun. Lett.}, vol.~28, no.~2, pp. 362--366, 2024.

\bibitem{sihlbom2022reconfigurable}
B.~Sihlbom, M.~I. Poulakis, and M.~{Di Renzo}, ``Reconfigurable intelligent surfaces: Performance assessment through a system-level simulator,'' \emph{IEEE Wireless Commun.}, vol.~30, no.~4, pp. 98--106, 2022.

\bibitem{yuan2024reconfigurable}
\BIBentryALTinterwordspacing
Y.~Yuan, Y.~Huang, X.~Su, B.~Duan, N.~Hu, and M.~{Di Renzo}, ``Reconfigurable intelligent surface ({RIS}) system level simulations for industry standards,'' 2024. [Online]. Available: \url{https://arxiv.org/abs/2409.13405}
\BIBentrySTDinterwordspacing

\bibitem{rafique2023reconfigurable}
A.~Rafique, N.~Ul~Hassan, M.~Zubair, I.~H. Naqvi, M.~Q. Mehmood, M.~{Di Renzo}, M.~Debbah, and C.~Yuen, ``Reconfigurable intelligent surfaces: Interplay of unit cell and surface-level design and performance under quantifiable benchmarks,'' \emph{IEEE Open Journal of the Communication Society}, vol.~4, pp. 1583--1599, 2023.

\bibitem{albanese2022marisa}
A.~Albanese, F.~Devoti, V.~Sciancalepore, M.~{Di Renzo}, and X.~Costa-Pérez, ``Marisa: A self-configuring metasurfaces absorption and reflection solution towards {6G},'' in \emph{IEEE INFOCOM 2022 - IEEE Conference on Computer Communications}, 2022, pp. 250--259.

\bibitem{9306896}
M.~Najafi, V.~Jamali, R.~Schober, and H.~V. Poor, ``Physics-based modeling and scalable optimization of large intelligent reflecting surfaces,'' \emph{IEEE Trans. Commun.}, vol.~69, no.~4, pp. 2673--2691, 2021.

\bibitem{Abrardo2021}
A.~Abrardo, D.~Dardari, and M.~{Di Renzo}, ``Intelligent reflecting surfaces: Sum-rate optimization based on statistical position information,'' \emph{IEEE Trans. Comm.}, vol.~69, no.~10, pp. 7121--7136, July 2021.

\bibitem{basar2019wireless}
E.~Basar, M.~{Di Renzo}, J.~De~Rosny, M.~Debbah, M.-S. Alouini, and R.~Zhang, ``Wireless communications through reconfigurable intelligent surfaces,'' \emph{IEEE Access}, vol.~7, pp. 116\,753--116\,773, 2019.

\bibitem{abrardo2021MIMO}
A.~Abrardo, D.~Dardari, M.~{Di Renzo}, and X.~Qian, ``{MIMO} interference channels assisted by reconfigurable intelligent surfaces: Mutual coupling aware sum-rate optimization based on a mutual impedance channel model,'' \emph{IEEE Wireless Commun. Lett.}, vol.~10, no.~12, pp. 2624--2628, 2021.

\bibitem{Wu2024Intelligent}
Q.~Wu, B.~Zheng, C.~You, L.~Zhu, K.~Shen, X.~Shao, W.~Mei, B.~Di, H.~Zhang, E.~Basar, L.~Song, M.~{Di Renzo}, Z.-Q. Luo, and R.~Zhang, ``Intelligent surfaces empowered wireless network: Recent advances and the road to {6G},'' \emph{Proc. IEEE}, pp. 1--40, 2024.

\bibitem{bartoli2023spatial}
G.~Bartoli, A.~Abrardo, N.~Decarli, D.~Dardari, and M.~{Di Renzo}, ``Spatial multiplexing in near field {MIMO} channels with reconfigurable intelligent surfaces,'' \emph{IET Signal Processing}, vol.~17, no.~3, p. e12195, 2023.

\bibitem{zander1992}
J.~Zander, ``Performance of optimum transmitter power control in cellular radio systems,'' \emph{IEEE Trans. Veh. Technol.}, vol.~41, no.~1, pp. 57--62, 1992.

\bibitem{yates1995}
R.~D. Yates, ``A framework for uplink power control in cellular radio systems,'' \emph{IEEE J. Sel. Areas Commun.}, vol.~13, no.~7, pp. 1341--1347, 1995.

\bibitem{bertsekas2016nonlinear}
D.~Bertsekas, \emph{Nonlinear Programming}.\hskip 1em plus 0.5em minus 0.4em\relax Athena Scientific, 2016, vol.~4.

\bibitem{5gchannelmodel2016}
{Aalto University and AT\&T and BUPT and CMCC and Ericsson and Huawei and INTEL and KT Corporation and Nokia and NTT DOCOMO and New York University and Qualcomm and Samsung and University of Bristol and University of Southern California}, ``{5G Channel Model for bands up to 100 GHz},'' Tech. Rep., October 2016.

\bibitem{palmucci2023two}
S.~Palmucci, A.~Guerra, A.~Abrardo, and D.~Dardari, ``Two-timescale joint precoding design and {RIS} optimization for user tracking in near-field {MIMO} systems,'' \emph{IEEE Trans. Signal Process.}, vol.~71, pp. 3067--3082, 2023.

\end{thebibliography}
\end{document}